\documentclass[longauth]{aa}
\usepackage[utf8]{inputenc}
\usepackage[varg]{txfonts}
\usepackage[breaklinks, colorlinks, citecolor=blue, linkcolor=blue]{hyperref}
\usepackage[usenames, dvipsnames]{color}
\usepackage{amsmath}
\usepackage{graphicx}
\usepackage[english]{babel}
\usepackage{siunitx}
\usepackage{float}
\usepackage{url}
\usepackage{longtable}
\usepackage{fancyhdr}
\usepackage{natbib}
\usepackage[normalem]{ulem}
\usepackage[version=4]{mhchem}

\makeatletter

\def\instrefs#1{{\def\scsep{\def\scsep{,}}\@for\w:=#1\do{\scsep\ref{inst:\w}}}}
\renewcommand{\inst}[1]{\unskip$^{\instrefs{#1}}$}

\let\orgautoref\autoref
\renewcommand{\autoref}
        {\def\equationautorefname{Eq.}%
         \def\figureautorefname{Fig.}%
         \def\sectionautorefname{Sect.}%
         \def\subsectionautorefname{Sect.}%
         \def\subsubsectionautorefname{Sect.}%
         \orgautoref}

\renewcommand*\aa@pageof{, page \thepage{} of \pageref*{LastPage}} 

\makeatother
\begin{document}

\title{Planetary system around the nearby M dwarf GJ~357 including a transiting, hot, Earth-sized planet optimal for atmospheric characterization
	\thanks{RV data are only available in electronic form
			at the CDS via anonymous ftp to cdsarc.u-strasbg.fr (130.79.128.5)
			or via \url{http://cdsweb.u-strasbg.fr/cgi-bin/qcat?J/A+A/}}
			}
\titlerunning{Planetary system around GJ~357}

\author{R.~Luque\inst{iac,ull} 
\and E.~Pall\'e\inst{iac,ull}
\and D.~Kossakowski\inst{mpia}
\and S.~Dreizler\inst{iag}
\and J.~Kemmer\inst{lsw}
\and N.~Espinoza\inst{mpia}
\and J.~Burt\inst{kavli,torres}
\and G.~Anglada-Escud\'e\inst{qm,iaa}
\and V.~J.~S.~B\'ejar\inst{iac,ull}
\and J.~A.~Caballero\inst{cabesac}
\and K.~A.~Collins\inst{cfa}
\and K.~I.~Collins\inst{vanderbilt}
\and M.~Cort\'es-Contreras\inst{cabesac}
\and E.~D\'iez-Alonso\inst{ucm,uovi}
\and F.~Feng\inst{dtm}
\and A.~Hatzes\inst{tls}
\and C.~Hellier\inst{keele}
\and T.~Henning\inst{mpia}
\and S.~V.~Jeffers\inst{iag}
\and L.~Kaltenegger\inst{cornell}
\and M.~K\"urster\inst{mpia}
\and J.~Madden\inst{cornell}
\and K.~Molaverdikhani\inst{mpia}
\and D.~Montes\inst{ucm}
\and N.~Narita\inst{iac,osa,jst,nao}
\and G.~Nowak\inst{iac,ull}
\and A.~Ofir\inst{wzm}
\and M.~Oshagh\inst{iag}
\and H.~Parviainen\inst{iac,ull}
\and A.~Quirrenbach\inst{lsw}
\and S.~Reffert\inst{lsw}
\and A.~Reiners\inst{iag}
\and C. Rodr\'iguez-L\'opez\inst{iaa}
\and M.~Schlecker\inst{mpia}
\and S.~Stock\inst{lsw}
\and T.~Trifonov\inst{mpia}
\and J.~N.~Winn\inst{princeton}
\and M.~R.~Zapatero Osorio\inst{cab}
\and M.~Zechmeister\inst{iag}
\and P.~J.~Amado\inst{iaa}
\and D.~R.~Anderson\inst{keele}
\and N.~E.~Batalha\inst{ucsc}
\and F.~F.~Bauer\inst{iaa}
\and P.~Bluhm\inst{lsw}
\and C.~J.~Burke\inst{kavli}
\and R.~P.~Butler\inst{dtm}
\and D.~A.~Caldwell\inst{seti,ames}
\and G.~Chen\inst{pmo}
\and J.~D.~Crane\inst{ocis}
\and D.~Dragomir\inst{kavli,hubble}
\and C.~D.~Dressing\inst{berkeley}
\and S.~Dynes\inst{kavli}
\and J.~M.~Jenkins\inst{ames}
\and A.~Kaminski\inst{lsw}
\and H.~Klahr\inst{mpia}
\and T.~Kotani\inst{osa,nao}
\and M.~Lafarga\inst{ice,ice2}
\and D.~W.~Latham\inst{cfa}
\and P.~Lewin\inst{lewin}
\and S.~McDermott\inst{protologic}
\and P.~Monta\~n\'es-Rodr\'iguez\inst{iac,ull}
\and J.~C.~Morales\inst{ice,ice2}
\and F.~Murgas\inst{iac,ull}
\and E.~Nagel\inst{hs}
\and S.~Pedraz\inst{caha}
\and I.~Ribas\inst{ice,ice2}
\and G.~R.~Ricker\inst{kavli}
\and P.~Rowden\inst{openac}
\and S.~Seager\inst{kavli,mit1,mit2}
\and S.~A.~Shectman\inst{ocis}
\and M.~Tamura\inst{osa,nao,ut}
\and J.~Teske\inst{ocis,hubble}
\and J.~D.~Twicken\inst{seti,ames}
\and R.~Vanderspeck\inst{kavli}
\and S.~X.~Wang\inst{dtm}
\and B.~Wohler\inst{seti,ames}
}

\institute{
\label{inst:iac}Instituto de Astrof\'isica de Canarias (IAC), 38205 La Laguna, Tenerife, Spain; \email{rluque@iac.es}
\and 
\label{inst:ull}Departamento de Astrof\'isica, Universidad de La Laguna (ULL), 38206, La Laguna, Tenerife, Spain
\and 
\label{inst:mpia}Max-Planck-Institut f\"ur Astronomie, K\"onigstuhl 17, 69117 Heidelberg, Germany
\and
\label{inst:iag}Institut f\"ur Astrophysik, Georg-August-Universit\"at, Friedrich-Hund-Platz 1, 37077 G\"ottingen, Germany
\and 
\label{inst:lsw}Landessternwarte, Zentrum f\"ur Astronomie der Universit\"at Heidelberg, K\"onigstuhl 12, 69117 Heidelberg, Germany
\and
\label{inst:kavli}Kavli Institute for Astrophysics and Space Research, Massachusetts Institute of Technology, Cambridge, MA 02139, USA
\and 
\label{inst:qm}School of Physics and Astronomy, Queen Mary, University of London, 327 Mile End Road, London, E1 4NS
\and 
\label{inst:iaa}Instituto de Astrof\'isica de Andaluc\'ia (IAA-CSIC), Glorieta de la Astronom\'ia s/n, 18008 Granada, Spain
\and 
\label{inst:cabesac}Centro de Astrobiolog\'ia (CSIC-INTA), ESAC, Camino bajo del castillo s/n, 28692 Villanueva de la Ca\~nada, Madrid, Spain
\and
\label{inst:cfa}Harvard-Smithsonian Center for Astrophysics, 60 Garden St, Cambridge, MA 02138, USA
\and
\label{inst:vanderbilt}Department of Physics and Astronomy, Vanderbilt University, Nashville, TN 37235, USA
\and 
\label{inst:ucm}Departamento de F\'isica de la Tierra y Astrof\'isica \& IPARCOS-UCM (Instituto de F\'isica de Part\'iculas y del Cosmos de la UCM), Facultad de Ciencias F\'isicas, Universidad Complutense de Madrid, 28040 Madrid, Spain
\and 
\label{inst:uovi}Department of Exploitation and Exploration of Mines, University of Oviedo, Oviedo, Spain
\and
\label{inst:dtm}Department of Terrestrial Magnetism, Carnegie Institution for Science, 5241 Broad Branch Road, NW, Washington, DC 20015, USA
\and 
\label{inst:tls}Th\"uringer Landessternwarte Tautenburg, Sternwarte 5, 07778 Tautenburg, Germany
\and
\label{inst:keele}Astrophysics Group, Keele University, Staffordshire, ST5 5BG, UK
\and
\label{inst:cornell}Carl Sagan Institute, Cornell University, Ithaca, NY 14853, USA
\and
\label{inst:osa}Astrobiology Center, 2-21-1 Osawa, Mitaka, Tokyo 181-8588, Japan
\and
\label{inst:jst}JST, PRESTO, 2-21-1 Osawa, Mitaka, Tokyo 181-8588, Japan
\and
\label{inst:nao}National Astronomical Observatory of Japan, 2-21-1 Osawa, Mitaka, Tokyo 181-8588, Japan
\and
\label{inst:wzm}Weizmann Institute of Science, 234 Herzl Street, Rehovot 761001, Israel
\and
\label{inst:princeton}Department of Astrophysical Sciences, Princeton University, 4 Ivy Lane, Princeton, NJ 08544, USA
\and 
\label{inst:cab}Centro de Astrobiolog\'ia (CSIC-INTA), Carretera de Ajalvir km 4, 28850 Torrej\'on de Ardoz, Madrid, Spain
\and
\label{inst:ucsc}Department of Astronomy and Astrophysics, University of California, Santa Cruz, CA 95064, USA
\and
\label{inst:seti}SETI Institute, Mountain View, CA 94043, USA
\and
\label{inst:ames}NASA Ames Research Center, Moffett Field, CA, 94035, USA
\and
\label{inst:pmo}Key Laboratory of Planetary Sciences, Purple Mountain Observatory, Chinese Academy of Sciences, Nanjing 210008, China
\and
\label{inst:ocis}Observatories of the Carnegie Institution for Science, 813 Santa Barbara Street, Pasadena, CA 91101, USA
\and 
\label{inst:berkeley}Department of Astronomy, The University of California, Berkeley, CA 94720, USA
\and
\label{inst:ice}Institut de Ci\`encies de l'Espai (ICE, CSIC), Campus UAB, C/Can Magrans s/n, 08193 Bellaterra, Spain
\and
\label{inst:ice2}Institut d’Estudis Espacials de Catalunya (IEEC), 08034 Barcelona,
Spain
\and
\label{inst:lewin}The Maury Lewin Astronomical Observatory, Glendora, California 91741, USA
\and
\label{inst:protologic}Proto-Logic LLC, 1718 Euclid Street NW, Washington, DC 20009, USA
\and 
\label{inst:hs}Hamburger Sternwarte, Universit\"at Hamburg, Gojenbergsweg 112, 21029 Hamburg, Germany
\and
\label{inst:caha}Centro Astron\'omico Hispano-Alem\'an (CSIC-MPG), Observatorio Astron\'omico de Calar Alto, Sierra de los Filabres-04550 G\'ergal, Almer\'ia, Spain
\and
\label{inst:openac}School of Physical Sciences, The Open University, Milton Keynes MK7 6AA, UK
\and
\label{inst:mit1}Department of Earth, Atmospheric and Planetary Sciences, Massachusetts Institute of Technology, Cambridge, MA 02139, USA
\and
\label{inst:mit2}Department of Aeronautics and Astronautics, MIT, 77 Massachusetts Avenue, Cambridge, MA 02139, USA
\and
\label{inst:ut}Department of Astronomy, The University of Tokyo, 7-3-1 Hongo, Bunkyo-ku, Tokyo 113-0033, Japan
\and
\label{inst:torres}Torres Fellow
\and
\label{inst:hubble}NASA Hubble Fellow
}

\date{Received 29 April 2019 / Accepted 27 June 2019}

\abstract{We report the detection of a transiting Earth-size planet around GJ~357, a nearby M2.5\,V star, using data from the Transiting Exoplanet Survey Satellite ({\it TESS}). GJ~357~b (TOI-562.01) is a transiting, hot, Earth-sized planet ($T_\mathrm{eq}=525\pm11\,\mathrm{K}$) with a radius of $R_\mathrm{b} = 1.217\pm0.084\,R_\oplus$ and an orbital period of $P_\mathrm{b} = 3.93$\,d. Precise stellar radial velocities from CARMENES and PFS, as well as archival data from HIRES, UVES, and HARPS also display a 3.93-day periodicity, confirming the planetary nature and leading to a planetary mass of $M_\mathrm{b} = 1.84\pm0.31\,M_\oplus$. In addition to the radial velocity signal for GJ~357~b, more periodicities are present in the data indicating the presence of two further planets in the system: GJ~357~c, with a minimum mass of $M_\mathrm{c} = 3.40\pm0.46\,M_\oplus$ in a 9.12\,d orbit, and GJ~357~d, with a minimum mass of $M_\mathrm{d} = 6.1\pm1.0\,M_\oplus$ in a 55.7\,d orbit inside the habitable zone. The host is relatively inactive and exhibits a photometric rotation period of $P_{\rm rot} = 78\pm2$\,d. GJ~357~b is to date the second closest transiting planet to the Sun, making it a prime target for further investigations such as transmission spectroscopy. Therefore, GJ~357~b represents one of the best terrestrial planets suitable for atmospheric characterization with the upcoming {\it JWST} and ground-based ELTs.}


\maketitle

\section{Introduction} \label{sec:intro}

To date nearly 200 exoplanets have been discovered orbiting approximately 100 M~dwarfs in the solar neighborhood \citep[e.g.,][]{2013A&A...549A.109B, Rowe2014ApJ...784...45R, Trifonov18, 2018Natur.563..365R}.  Some of these orbit near to or in the habitable zone \citep[e.g.,][]{2007A&A...469L..43U, 2013A&A...556A.126A, 2016Natur.536..437A, Tuomi2013A&A...556A.111T, 2017Natur.544..333D, Reiners17}.  However, only 11 M~dwarf planet systems have been detected with both the transit as well as the radial velocity (RV) method, which allows us to derive their density from their measured radius and mass, informing us about its bulk properties. When transit timing variation (TTV) mass measurements are included, TRAPPIST-1 \citep[2MUCD~12171,][]{Gillon2017Natur.542..456G} represents the 12th M~dwarf planet system with mass and radius measurements.

Only six of the abovementioned eleven systems contain planets with masses below $10\,M_\oplus$:
LHS~1140~b and~c \citep[GJ~3053,][]{2017Natur.544..333D, 2019AJ....157...32M},
K2-3~b and~c \citep[PM~J11293--0127,][]{Almenara2015A&A...581L...7A, Sinukoff2016ApJ...827...78S},
K2-18~b \citep[PM J11302+0735,][]{Cloutier2017A&A...608A..35C, Sarkis2018AJ....155..257S},
GJ~1214~b \citep[LHS~3275~b,][]{Harpsoe2013A&A...549A..10H}, GJ~1132 \citep{GJ1132,Bonfils2018AA...618A.142B}, and 
b--g planets of TRAPPIST-1 \citep{Gillon2016Natur.533..221G, Gillon2017Natur.542..456G}.  
However, only three planets with masses similar to Earth orbit M~dwarfs of moderate brightness ($J$ = 9.2--9.8\,mag): GJ~1132~b ($1.66 \pm 0.23\,M_\oplus$), LHS~1140~c ($1.81\pm0.39\,M_\oplus$), and K2-18~b (2.1$^{+2.1}_{-1.3}\,M_\oplus$). Systems hosting small terrestrial exoplanets orbiting bright stars are ideal not only from the perspective of precise mass measurements with ground-based instruments, but also for further orbital (e.g., obliquity determination) and atmospheric characterization using current and future observatories \citep[see, e.g., ][]{batalha:2018}.

The Transiting Exoplanet Survey Satellite \citep[{\it TESS},][]{Ricker2015} mission is an observatory that was launched to find small planets transiting small, bright stars. Indeed, since the start of scientific operations in July 2018, {\it TESS} has already uncovered over 600 new planet candidates, and is quickly increasing the sample of known Earths and super-Earths around small M-type stars \citep{Vanderspek2019ApJ...871L..24V, Guenther2019arXiv190306107G, 2019arXiv190308017K}. In this paper, we present the discovery of three small planets around a bright M dwarf, one of which,  GJ~357~b, is an Earth-sized transiting exoplanet discovered using photometry from the {\it TESS} mission. To date, GJ~357~b is the second nearest ($d=9.44\,\mathrm{pc}$) transiting planet to the Sun after HD~219134~b \citep[][$d=6.53\,\mathrm{pc}$]{Motalebi2015A&A...584A..72M}, and the closest around an M dwarf. Besides, it is amenable to future detailed atmospheric characterization, opening the door to new studies for atmospheric characterization of Earth-like planet atmospheres \citep{Palle2009Natur.459..814P}.

The paper is structured as follows. Section \ref{sec:tess} presents the {\it TESS} photometry used for the discovery of GJ~357~b. Section \ref{sec:obs} presents ground-based observations of the star including seeing-limited photometric monitoring, high-resolution imaging, and precise RVs. Section \ref{subsec:star} presents a detailed analysis of the stellar properties of GJ~357. Section \ref{sec:fit} presents an analysis of the available data in order to constrain the planetary properties of the system, including precise mass constraints on GJ~357~b along with a detection and characterization of two additional planets in the system, GJ~357~c and GJ~357~d. Section \ref{sec:discussion} presents a discussion of our results and, finally, Sect. \ref{sec:conclusions} presents our conclusions.

\section{\textit{TESS} photometry} \label{sec:tess}

\begin{figure*}[ht!]
    \centering
    \includegraphics[width=\hsize]{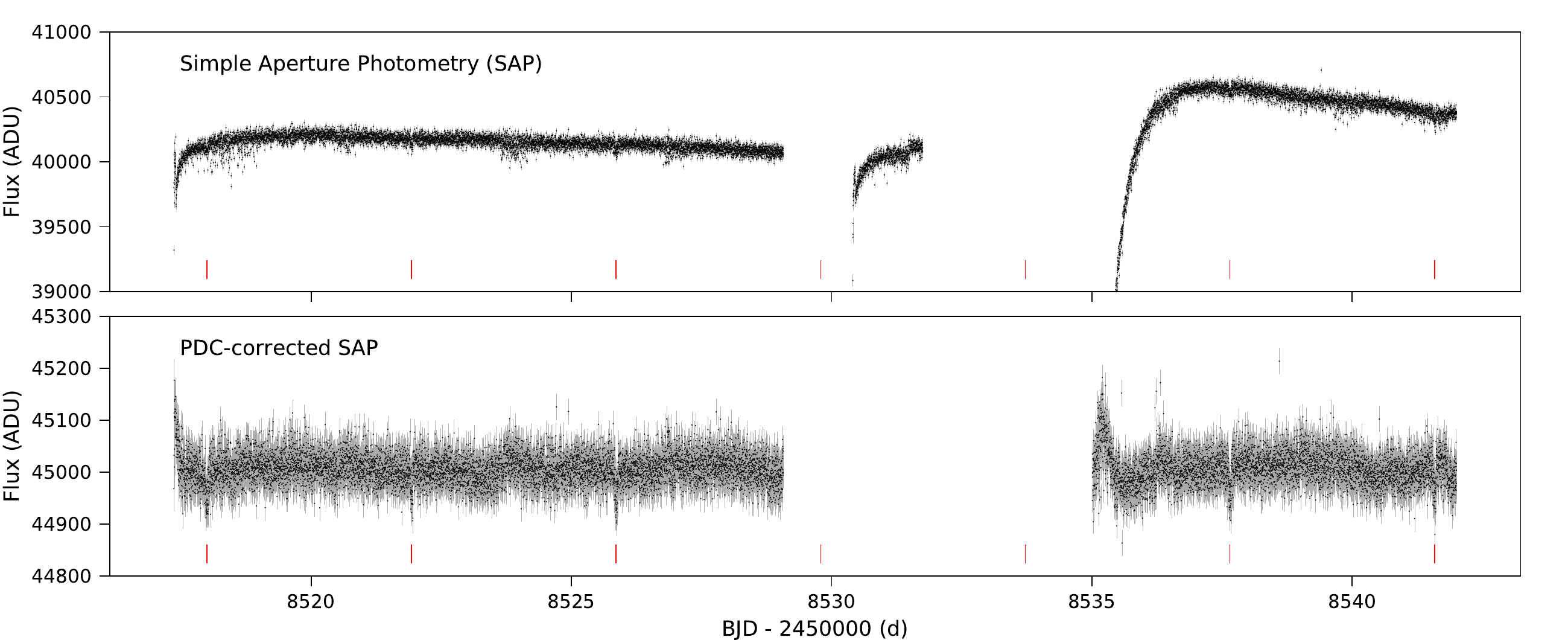}
    \caption{{\it TESS} light curves of GJ~357 provided by SPOC. Top panel: Simple aperture photometry. Bottom panel: PDC-corrected photometry. Transits of the planet candidate TOI-562.01 are marked in red.}
    \label{fig:tess_lc}
\end{figure*}

Planet GJ~357 (TIC 413248763) was observed by \textit{TESS} in 2-min short-cadence integrations in Sector 8 (Camera \#2, CCD \#3) from February 2, 2019 until February 27, 2019 (see \autoref{fig:tess_lc}), and will not be observed again during the primary mission. At $\mathrm{BJD}=2458531.74$, an interruption in communications between the instrument and spacecraft occurred, resulting in an instrument turn-off until $\mathrm{BJD}=2458535.00$. Together with the satellite repointing for data downlink between $\mathrm{BJD}=2458529.06$ and $\mathrm{BJD}=2458530.44$, a gap of approximately 6\,d is present in the photometry. In our analysis, the datapoints between $\mathrm{BJD}=2458530.44$ and $\mathrm{BJD}=2458531.74$ were masked out.

\subsection{Transit searches} \label{subsec:transit}

\textit{TESS} objects of interest (TOIs) are announced regularly via the \textit{TESS} data alerts public website\footnote{\url{https://tess.mit.edu/alerts/}}. TOI-562.01 was announced on April 13, 2019 and its corresponding light curve produced by the Science Processing Operations Center \citep[SPOC;][]{SPOC} at the NASA Ames Research Center was uploaded to the Mikulski Archive for Space Telescopes (MAST)\footnote{\url{https://mast.stsci.edu}} on April 17, 2019. SPOC provided for this target simple aperture photometry (SAP) and systematics-corrected photometry, a procedure consisting of an adaptation of the Kepler Presearch Data Conditioning algorithm \citep[PDC,][]{Smith2012PASP..124.1000S, Stumpe2012PASP..124..985S, Stumpe2014PASP..126..100S} to \textit{TESS}. The light curves generated by both methods are shown in \autoref{fig:tess_lc}. We use the latter one (PDC-corrected SAP, \autoref{fig:tess_lc} bottom panel) for the remainder of this work.

A signal with a period of 3.93\,d and a transit depth of $1164 \pm 66\,\mathrm{ppm}$, corresponding to a planet radius of approximately $1.3\pm0.3\,R_\oplus$ was detected in the {\it TESS} photometry. The Earth-sized planet candidate passed all the tests from the Alerts Data Validation Report\footnote{The complete DVR of TOI-562.01 can be downloaded from \url{https://tev.mit.edu/vet/spoc-s08-b01/413248763/dl/pdf/}.} \citep[DVR;][]{Twicken2018PASP..130f4502T, Li2019PASP..131b4506L}, for example, even-odd transits comparison, eclipsing binary discrimination tests, ghost diagnostic tests to help rule out scattered light, or background eclipsing binaries, among others. The report indicates that the dimming events are associated with significant image motion, which is usually indicative of a background eclipsing binary. However, in this case, the reported information is meaningless because the star is saturated. On the other hand, the transit source is coincident with the core of the stellar point spread function (PSF), so the transit events happen on the target and not, for example, on a nearby bright star. 

We also performed  an independent analysis of the \textit{TESS} light curve in order to confirm the DVR analysis and search for additional transit signals. An iterative approach was employed: in each iteration the same raw data were detrended and outliers-rejected, a signal was identified and then modeled, and that model was temporarily divided-out during the detrending of the next iteration to produce a succession of improving models --- until the $\chi^2$ converges. The raw photometry was detrended by fitting a truncated Fourier series, starting from the natural period of twice the data span, and all of its harmonics, down to some "protected" time span to make sure the filter does not modify the shape of the transit itself. We used a protected time span of 0.5\,d, and this series was iteratively fitted with 4$\sigma$ rejection. Finally, \texttt{OptimalBLS} \citep{Ofir14} is used to identify the transit signal, which is then modeled using the \citet{MandelAgol} model and the differential evolution Markov Chain Monte Carlo algorithm \citep{terBraak2008}. The final model has $\chi^2_\nu=1.017$ and the resultant transit parameters are consistent with the \textit{TESS} DVR. We also checked for odd-even differences between the transits, additional transit signals, and parabolic TTVs \citep{Ofir18} -- all with null results.

\subsection{Limits on photometric contamination} \label{subsec:contamination}

Given the large \textit{TESS} pixel size of 21\arcsec, it is essential to verify that no visually close-by targets are present that could affect the depth of the transit. There are two bright objects within one \textit{TESS} pixel of GJ~357: ($i$) \textit{Gaia} DR2 5664813824769090944 at 15.19\arcsec\ and $G_{R_P} = 15.57$\,mag; and ($ii$) \textit{Gaia} DR2 5664814202726212224 at 18.31\arcsec\ and $G_{R_P} = 15.50\,\mathrm{mag}$. However, they are much fainter than GJ~357 (TOI-562, \textit{Gaia} DR2 5664814198431308288, $G_{R_P}=8.79$\,mag) and their angular separations actually increased between the epochs of observation of \textit{Gaia} (J2015.5) and \textit{TESS} (J2019.1--J2019.2) due to the high proper motion of this star. 

These two sources are by far the brightest ones apart from our target in the digitizations of red photographic plates taken in 1984 and 1996 with the UK Schmidt telescope. The \textit{Gaia} $G_{R_P}$-band (630--1050\,nm) and the \textit{TESS} band (600--1000\,nm) are very much alike, allowing us to estimate the dilution factor for \textit{TESS} using Eq.~2 in \citet{juliet} to be $D_{\it TESS}=0.996$, which is consistent with 1.00, therefore compatible with no flux contamination.

\section{Ground-based observations} \label{sec:obs}

\subsection{Transit follow-up}

We acquired ground-based time-series follow-up photometry of a full transit of TOI-562.01 on UTC April 26, 2019 from a Las Cumbres Observatory (LCO) 1.0\,m telescope \citep{Brown:2013} at Cerro Tololo Inter-American Observatory (CTIO) as part of the {\it TESS} follow-up program (TFOP) SG1 Group. We used the {\tt TESS Transit Finder}, which is a customized version of the {\tt Tapir} software package \citep{Jensen:2013}, to schedule photometric time-series follow-up observations. The $4096\times4096$ LCO SINISTRO camera has an image scale of 0$\farcs$389 pix$^{-1}$ resulting in a $26\arcmin\times26\arcmin$ field of view. The 227\,min observation in $z_s$ band used 30\,s exposure times which, in combination with the 26\,s readout time, resulted in 244 images. The images were calibrated by the standard LCO {\tt BANZAI} pipeline and the photometric data were extracted using the {\tt AstroImageJ} software package \citep{Collins:2017}. The target star light curve shows a clear transit detection in a 7.78\arcsec\ radius aperture (see middle right panel of Fig.~\ref{fig:jointfit}). The full width half maximum (FWHM) of the target and nearby stars is $\sim$ 4\arcsec, so the follow-up aperture is only marginally contaminated by neighboring faint {\it Gaia} DR2 stars. The transit signal can be reliably detected with apertures that have a radius as small as 4.28\arcsec, after which systematic effects start to dominate the light curve. We note that the detection of an $\sim 1200\,\mathrm{ppm}$ transit with a 1\,m ground-based telescope in a single transit is remarkable. A similar performance has been achieved only with the 1.2\,m Euler-Swiss telescope combining two transits of HD~106315~c \citep{Lendl2017A&A...603L...5L}, and highlights the importance of ground-based facilities to maintain and refine ephemeris of {\it TESS} planet candidates even in the Earth-sized regime.

\subsection{Seeing-limited photometric monitoring} \label{subsec:phot}

We made a compilation of photometric series obtained by long-time baseline, automated surveys exactly as in \mbox{\citet{DiezAlonso2019A&A...621A.126D}}. In particular we retrieved data from the following public surveys: All-Sky Automated Survey \citep[ASAS;][]{Pojmanski2002AcA....52..397P}, Northern Sky Variability Survey \citep[NSVS;][]{Wozniak2004AJ....127.2436W}, and All-Sky Automated Survey for Supernovae \citep[ASAS-SN;][]{Kochanek2017PASP..129j4502K}. The telescope location, instrument configurations, and photometric bands of each public survey were summarized by \cite{DiezAlonso2019A&A...621A.126D}. We did not find GJ~357 data in other public catalogs, such as The MEarth Project \citep{MEarth}, the Catalina surveys \citep{Drake2014ApJS..213....9D}, or the Hungarian Automated Telescope Network \citep{Bakos2004PASP..116..266B}.

WASP-South, the southern station of the Wide Angle Search for Planets  \citep{Pollacco2006PASP..118.1407P}, is an array of eight cameras using 200-mm f/1.8 lenses backed by $2048\times2048$ CCDs, each camera covering $7.8^{\circ} \times 7.8^{\circ}$. It rasters a set of different pointings with a typical 10-min cadence.  WASP-South observed fields containing GJ~357 every year from 2007 to 2012, obtaining data over a span of typically 120\,d each season, acquiring a total of 48\,000 photometric observations.

\subsection{High-resolution imaging} \label{subsec:fastcam}

\begin{figure}
    \centering
    \includegraphics[width=\hsize]{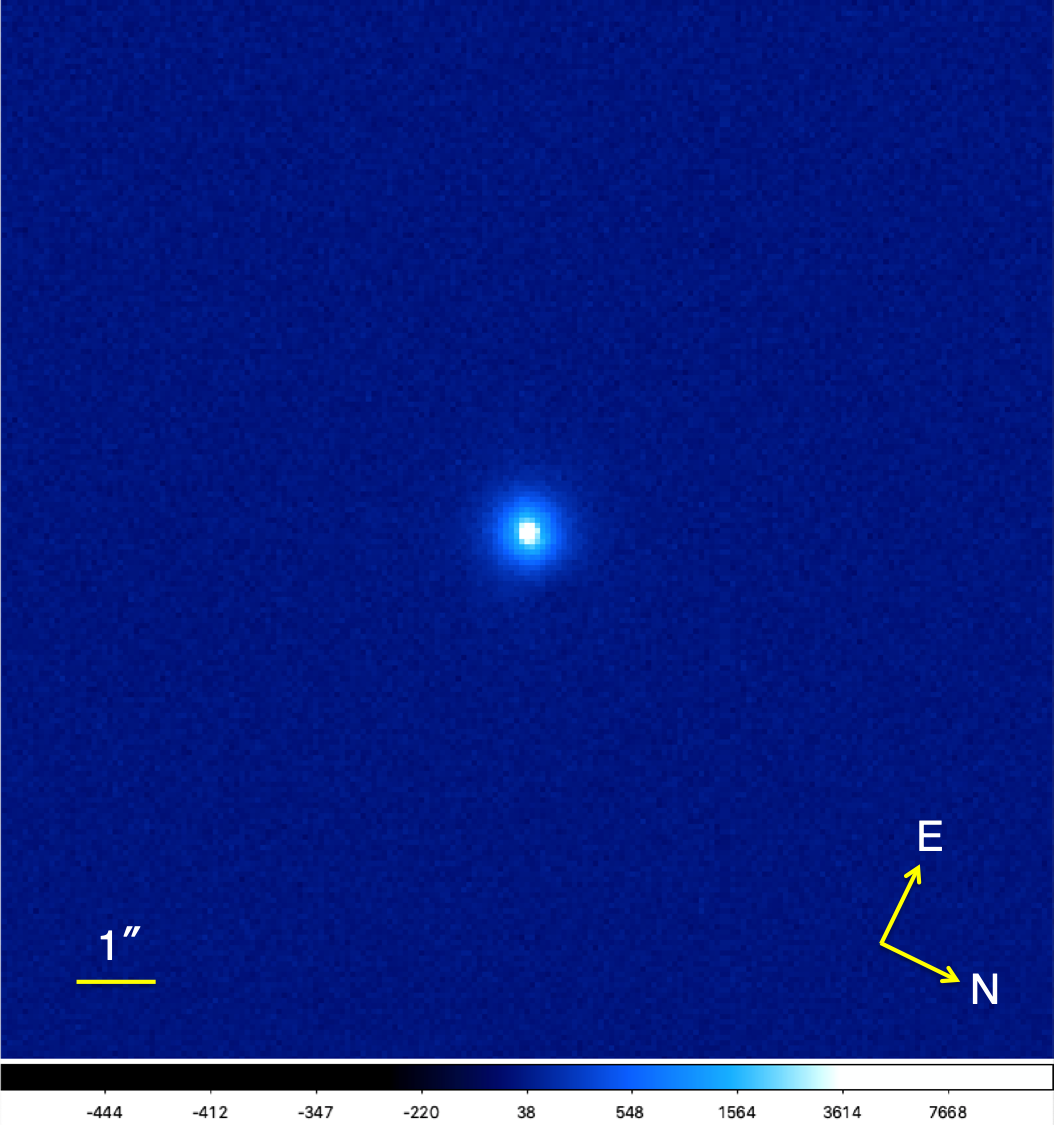}
    \caption{Adaptive-optics-corrected image taken with the fiber injection module camera of IRD mounted on the Subaru 8.2\,m telescope on April 18, 2019. The field of view is $13.4\arcsec\times13.4\arcsec$ ($200\times200$\,pix). Color coding is assigned in logarithmic scale. North is 116\,deg turned clockwise from the upper direction.
    }
    \label{fig:hri}
\end{figure}

\paragraph{FastCam}
Although we discuss in Sect.~\ref{subsec:contamination} that there are no visually close companions that could affect the depth of the transit of GJ~357, we obtained high-resolution observations at different epochs to exclude the possibility of a physically-bound eclipsing binary that may produce the transits detected in the {\it TESS} light curve. First, we observed GJ~357 with the FastCam instrument \citep{2008SPIE.7014E..47O} mounted on the 1.5\,m Telescopio Carlos S\'anchez at the Teide Observatory on January 14, 2013. These observations were part of our high-resolution imaging campaign of M~dwarfs to characterize stellar multiplicity and select the most appropriate targets for the CARMENES survey \citep{CC17}. FastCam is a lucky imaging camera with a high readout speed, employing the subelectron noise L3CCD Andor $512\times512$ detector, which provides a pixel size of 0.0425\arcsec\ and a field of view of $21.2\arcsec \times 21.2\arcsec$. We obtained ten blocks of a thousand individual frames with 50\,ms exposure time in the $I$ band. Data were bias subtracted, aligned, and combined using the brightest pixel as a reference as described in  \citet{2010SPIE.7735E..0XL} and \citet{2013MNRAS.429..859J}. We selected the best 10\,\% of the frames to produce the final image and determined that there are no background contaminating sources with $\delta\,I<3\,\mathrm{mag}$ down to 0.5\arcsec\ and with  $\delta\,I<6\,\mathrm{mag}$ down to 3.0\arcsec\ and up to 8.5\arcsec\ (given by the detector size). 

\paragraph{IRD}
We also observed GJ~357 with the InfraRed Doppler \citep[IRD,][]{Kotani2018SPIE10702E..11K} instrument on the Subaru 8.2\,m telescope on April 18, 2019. IRD is a fibre-fed instrument through a fibre injection module behind an adaptive optics (AO) system \citep[AO188,][]{Hayano2010SPIE.7736E..0NH}. A fibre injection module camera (FIMC) monitors images around targets to enable fibre injection of stellar light and guiding, and can take AO-corrected images of observing targets. The FIMC employs a CCD with pixel scale of 0.067\arcsec{} per pixel and observes in 970--1050\,nm. Figure~\ref{fig:hri} is the FIMC image of GJ~357. The image shows a $200\times200$ pixel region around GJ~357, revealing no nearby point source. 

We note that GJ~357 is a high proper motion star with 0.139\arcsec\ per year in RA and -0.990\arcsec{} per year in DEC based on the {\it Gaia} Data Release 2 (DR2) data \citep{GaiaDR2}. It means that the star was about 0.8\arcsec{} to the west and about 6\arcsec{} to the north at the time of the FastCam observation. The FastCam and IRD FIMC non-detection of any nearby companion excludes any background object at the original position of the FastCam observation. We exclude any false positive scenario and conclude there is no flux contamination from visually close-by targets in the GJ~357 transit data, and we fix the dilution factor for {\it TESS} to one in all of our model fits.


\subsection{Precise radial velocities}

\subsubsection{HIRES}

The high-resolution spectrograph HIRES \citep{HIRES} mounted on the 10-m Keck-I telescope has been extensively used to search for exoplanets around bright dwarf stars using the RV technique \citep[e.g.,][]{2000ApJ...536..902V,2008PASP..120..531C}.  As part of this effort, \citet{Butler17} published 64\,480 observations of a sample of 1699 stars collected with HIRES between 1996 and 2014. These data have been recently reanalyzed by \citet{Tal-Or19} using a sample of RV-quiet stars (i.e., whose RV scatter is $< 10\,\mathrm{m\,s^{-1}}$),  who found small, but significant systematic effects in the RVs: a discontinuous jump caused by major modifications of the instrument in August 2004, a long-term drift, and a small intra-night drift. We use a total of 36 measurements for GJ~357 taken between January 26, 1998 and February 20, 2013. The RVs show a median internal uncertainty of $2.4\,\mathrm{m\,s^{-1}}$ and a $\mathrm{rms}$ of $4.0\,\mathrm{m\,s^{-1}}$ around the mean value.

\subsubsection{UVES}

\citet{Zechmeister2009A&A...505..859Z} published 70 RV measurements of GJ~357 taken between November 15, 2000 and March 25, 2007 as part of the M~dwarf planet search for terrestrial planets in the habitable zone with UVES at the ESO Very Large Telescope. RVs were obtained with the \texttt{AUSTRAL} code \citep{2000A&A...362..585E} and combined into nightly averages following \citet{2003A&A...403.1077K}.  The 30 nightly binned RVs show a median internal uncertainty of $2.5\,\mathrm{m\,s^{-1}}$ and a $\mathrm{rms}$ of $5.3\,\mathrm{m\,s^{-1}}$ around the mean value.

\subsubsection{HARPS}

The High Accuracy Radial velocity Planet Searcher \citep[HARPS,][]{HARPS} is an ultra-precise \'Echelle spectrograph in the optical regime installed at the ESO 3.6\,m telescope at La Silla Observatory in Chile, with a sub-$\mathrm{m\,s^{-1}}$ precision.  We retrieved 53 high-resolution spectra from the ESO public archive collected between December 13, 2003 and February 13, 2013. We extracted the FWHM and bisector span (BIS) of the cross-correlation function from the FITS headers as computed by the DRS ESO HARPS pipeline \citep{2007A&A...468.1115L}, but to obtain the RVs we used SERVAL \citep{SERVAL}, based on least-squares fitting with a high signal-to-noise (S/N) template created by co-adding all available spectra of the star.  The RVs have a median internal uncertainty of $1\,\mathrm{m\,s^{-1}}$ and a $\mathrm{rms}$ of $3.3\,\mathrm{m\,s^{-1}}$ around the mean value.

\subsubsection{PFS}

The Planet Finder Spectrograph \citep{Crane2010} is an iodine-cell, high-precision RV instrument mounted on the 6.5\,m Magellan II telescope at Las Campanas Observatory in Chile. RVs are measured by placing a cell of gaseous I$_2$ in the converging beam of the telescope. This imprints the 5000-6200\,\AA\ region of incoming stellar spectra with a dense forest of I$_2$ lines that act as a wavelength calibrator, and provide a proxy for the point spread function (PSF) of the instrument. GJ~357 was observed a total of nine times as part of the long-term Magellan Planet Search Program between March 2016 and January 2019. After {\it TESS}' identification of transits in GJ~357, the star was then observed at higher precision during the April and May 2019 runs, which added an additional seven RVs to the dataset. The iodine data prior to February 2018 (PFSpre) were taken through a 0.5" slit resulting in $R\sim 80,000$, and those after (PFSpost) were taken through a 0.3" slit, resulting in $R\sim 130,000$. A different offset must be accounted for the RVs taken before and after this intervention. All PFS data are reduced with a custom IDL pipeline that flat fields, removes cosmic rays, and subtracts scattered light. Additional details about the iodine-cell RV extraction method can be found in \citet{Butler1996}. The RVs have a median internal uncertainty of $1.3~(0.7)\,\mathrm{m\,s^{-1}}$ and an $\mathrm{rms}$ of $3.1~(2.3)\,\mathrm{m\,s^{-1}}$ around the mean value for PFSpre (PFSpost).   

\subsubsection{CARMENES}

The star GJ~357 (Karmn J09360-216) is one of the 342 stars monitored in the CARMENES Guaranteed Time Observation program to search for exoplanets around M dwarfs, which began in January 2016 \citep{Reiners17}. The CARMENES instrument is mounted at the 3.5\,m telescope at the Calar Alto Observatory in Spain and has two channels: the visual (VIS) covers the spectral range $0.52$--$\SI{0.96}{\micro\metre}$ and the near-infrared (NIR) covers the $0.96$--$\SI{1.71}{\micro\metre}$ range \citep{CARMENES, CARMENES18}.
GJ~357 was observed ten times between December 13, 2016 and March 16, 2019, and the VIS RVs -- extracted with SERVAL and corrected for barycentric motion, secular acceleration, instrumental drift, and nightly zero-points \citep[see][for details]{Trifonov18, Luque18} -- show a median internal uncertainty of $1.3\,\mathrm{m\,s^{-1}}$ and a $\mathrm{rms}$ of $2.8\,\mathrm{m\,s^{-1}}$ around the mean value.

\section{Stellar properties} \label{subsec:star}

\begin{table}
\centering
\small
\caption{Stellar parameters of GJ~357.} \label{tab:star}
\begin{tabular}{lcr}
\hline\hline
\noalign{\smallskip}
Parameter                               & Value                 & Reference \\ 
\hline
\noalign{\smallskip}
\multicolumn{3}{c}{Name and identifiers}\\
\noalign{\smallskip}
Name                            & L 678-39                      & {\citet{1942POMin...2..242L}}      \\
GJ                              & 357                           & {\citet{1957MiABA...8....1G}}      \\
Karmn                           & J09360-216                    & {AF15}      \\    
TOI                             & 562                           & {\it TESS} Alerts      \\  
TIC                             & 413248763                     & {\citet{2018AJ....156..102S}}      \\  
\noalign{\smallskip}
\multicolumn{3}{c}{Coordinates and spectral type}\\
\noalign{\smallskip}
$\alpha$                                & 09:36:01.64           & {\it Gaia} DR2     \\
$\delta$                                & --21:39:38.9           & {\it Gaia} DR2     \\
SpT                                     & M2.5\,V                & {\citet{PMSU}}             \\
\noalign{\smallskip}
\multicolumn{3}{c}{Magnitudes}\\
\noalign{\smallskip}
$B$ [mag]                               & $12.52\pm0.02$        & UCAC4       \\
$V$ [mag]                               & $10.92\pm0.03$        & UCAC4       \\
$g$ [mag]                               & $11.70\pm0.02$        & UCAC4       \\
$G$ [mag]                               & $9.8804\pm0.0014$        & {\it Gaia} DR2       \\
$r$ [mag]                               & $10.34\pm0.09$        & UCAC4       \\
$i$ [mag]                               & $9.35\pm0.27$         & UCAC4       \\
$J$ [mag]                               & $7.337\pm0.034$       & 2MASS       \\
$H$ [mag]                               & $6.740\pm0.033$       & 2MASS       \\
$K_s$ [mag]                             & $6.475\pm0.017$       & 2MASS       \\
\noalign{\smallskip}
\multicolumn{3}{c}{Parallax and kinematics}\\
\noalign{\smallskip}
$\pi$ [mas]                             & $105.88\pm0.06$       & {\it Gaia} DR2             \\
$d$ [pc]                                & $9.444\pm0.005 $      & {\it Gaia} DR2             \\
$\mu_{\alpha}\cos\delta$ [$\mathrm{mas\,yr^{-1}}$]  & $+138.694 \pm 0.100$ & {\it Gaia} DR2          \\
$\mu_{\delta}$ [$\mathrm{mas\,yr^{-1}}$]            & $-990.311 \pm 0.083$ & {\it Gaia} DR2          \\
$V_r$ [$\mathrm{km\,s^{-1}}]$           & -34.70$\pm$0.50    & This work    \\
$U$ [$\mathrm{km\,s^{-1}}]$             & 41.11$\pm$0.13    & This work      \\
$V$ [$\mathrm{km\,s^{-1}}]$             & 11.37$\pm$0.45    & {This work}      \\
$W$ [$\mathrm{km\,s^{-1}}]$             & -37.25$\pm$0.19    & {This work}      \\
\noalign{\smallskip}
\multicolumn{3}{c}{Photospheric parameters}\\
\noalign{\smallskip}
$T_{\mathrm{eff}}$ [K]                      & $3505 \pm 51$         & \citet{2019arXiv190403231S}   \\
$\log g$                                    & $4.94 \pm 0.07$       & \citet{2019arXiv190403231S}   \\
{[Fe/H]}                                    & $-0.12 \pm 0.16$      & \citet{2019arXiv190403231S}   \\
$v \sin i_\star$ [$\mathrm{km\,s^{-1}}$]    & $<2.0$                & \citet{Reiners17}             \\
\noalign{\smallskip}
\multicolumn{3}{c}{Physical parameters}\\
\noalign{\smallskip}
$M$ [$M_{\odot}$]                       & $0.342 \pm 0.011$     & \citet{2019arXiv190403231S}       \\
$R$ [$R_{\odot}$]                       & $0.337 \pm 0.015$     & \citet{2019arXiv190403231S}       \\
$L$ [$10^{-4}\,L_\odot$]                & $159.1 \pm 3.6$       & \citet{2019arXiv190403231S}       \\
\noalign{\smallskip}
\hline
\end{tabular}
\tablebib{
    AF15: \citet{2015A&A...577A.128A};
    {\it Gaia} DR2: \citet{GaiaDR2};
    UCAC4: \citet{UCAC4};
    2MASS: \citet{2MASS}.
}
\end{table}

\subsection{Stellar parameters}

The star GJ~357 (L~678-39, Karmn~J09360-216, TIC~413248763) is a high proper motion star in the Hydra constellation classified as M2.5\,V by \citet{PMSU}.  Located at a distance of $d \approx 9.4\,\mathrm{pc}$ \citep{GaiaDR2}, it is one of the brightest single M~dwarfs in the sky, with an apparent magnitude in the $J$ band of 7.337\,mag \citep{2MASS} and no evidence for multiplicity, either at short or wide separations \citep{CC17}. Accurate stellar parameters of GJ~357 were presented in \citet{2019arXiv190403231S}, who determined radii, masses, and updated photospheric parameters for 293 bright M~dwarfs from the CARMENES survey using various methods. In summary, \citet{2019arXiv190403231S} derived the radii from Stefan-Boltzmann's law, effective temperatures from a spectral analysis using the latest grid of PHOENIX-ACES models, luminosities from integrating broadband photometry together with {\it Gaia} DR2 parallaxes, and masses from an updated mass-radius relation derived from eclipsing binaries.

According to this analysis, GJ~357 has an effective temperature of $3505 \pm 51\,\mathrm{K}$ and a mass of $0.342 \pm 0.011\,M_{\odot}$. Furthermore, with the {\it Gaia} DR2 equatorial coordinates, proper motions, and parallax, and absolute RV measured from CARMENES spectra, we compute galactocentric space velocities $UVW$ as in \cite{2001MNRAS.328...45M} and \cite{Cortes2016UCM-PhD} that kinematically put GJ~357 in the thin disk of the Galaxy. A summary of all stellar properties can be found in Table~\ref{tab:star}.

\subsection{Activity and rotation period} \label{subsec:rotation}

Using CARMENES data, \citet{Reiners17} determined a Doppler broadening upper limit of $v \sin i < 2\,\mathrm{km\,s^{-1}}$ for GJ~357. This slow rotational velocity is consistent with its low level of magnetic activity. An analysis of the H$\alpha$ activity in the CARMENES spectra shows that it is an inactive star and that the rotational variations in H$\alpha$ and other spectral indicators are consistent with other inactive stars \citep{Schoefer19}. GJ~357 has a $\log R'_{HK}$ value of --5.37, and is one of the least active stars in the \citet{2018A&A...616A.108B} catalog of chromospheric activity of nearly 4500 stars, consistent with our kinematic analysis. In addition, this is in agreement with the upper limit set by \citet{2013MNRAS.431.2063S} in its X-ray flux ($\log F_X < -13.09\,\mathrm{mW\,m^{-2}}$) and the fact that \citet{Moutou17} were not able to measure its magnetic field strength based on optical high-resolution spectra obtained with ESPaDOnS at the Canada-France-Hawai'i Telescope.

From spectroscopic determinations, the small value of $\log R'_{HK}$ indicates a long rotation period of between 70 and 120\,d \citep{2015MNRAS.452.2745S,2017A&A...600A..13A,2018A&A...616A.108B}. Therefore, we searched the WASP data for rotational modulations, treating each season of data in a given camera as a separate dataset, using the methods presented in \citet{Maxted2011PASP..123..547M}. The results are tabulated in Table~\ref{tab:phot}. We find a significant 70--90\,d periodicity across all seasons with more than 2000 datapoints. Since this timescale is not much shorter than the coverage in each year, the period error in each dataset is $\sim10$\,d. The amplitude of the modulation ranges from 2 to 9\,mmag, and the false-alarm probability is less than 10$^{-4}$. 

\begin{table}
    \centering
    \small
    \caption{Rotation-modulation search of WASP-South data.} \label{tab:phot}
    \begin{tabular}{lccccc}
        \hline\hline
        \noalign{\smallskip}
   Year (camera)   &  $N_{\rm pts}$   & $P$    & Ampl.         &  FAP   &  A$_{95}$\tablefootmark{a}    \\
          &        & (d)    & (mag)         &        &  (mag)  \\
        \noalign{\smallskip}
        \hline
        \noalign{\smallskip}
2007 (226)  &  7225  &   74  &   0.002  &  0.0099  &  0.0011  \\
2008 (226)  &  8947  &   79  &   0.002  &  0.0083  &  0.0013  \\
2009 (228)  &  5785  &   91  &   0.005  &  0.0000  &  0.0015  \\
2010 (227)  &  2291  &   84  &   0.006  &  0.0001  &  0.0036 \\
2010 (228)  &  4745  &   84  &   0.009  &  0.0000  &  0.0038  \\
2011 (222)  &  5272  &   72  &   0.004  &  0.0000  &  0.0019  \\
2012 (222)  &  5085  &   74  &   0.007  &  0.0000  &  0.0026  \\
2012 (227)  &  2605  &   71  &   0.006  &  0.0000  &  0.0030  \\
        \noalign{\smallskip}         
        \hline
    \end{tabular}
    \tablefoot{
        \tablefoottext{a}{Amplitude corresponding to a 95\% probability of a false alarm.}
    }
\end{table}

\begin{figure}
    \centering
    \includegraphics[width=\hsize]{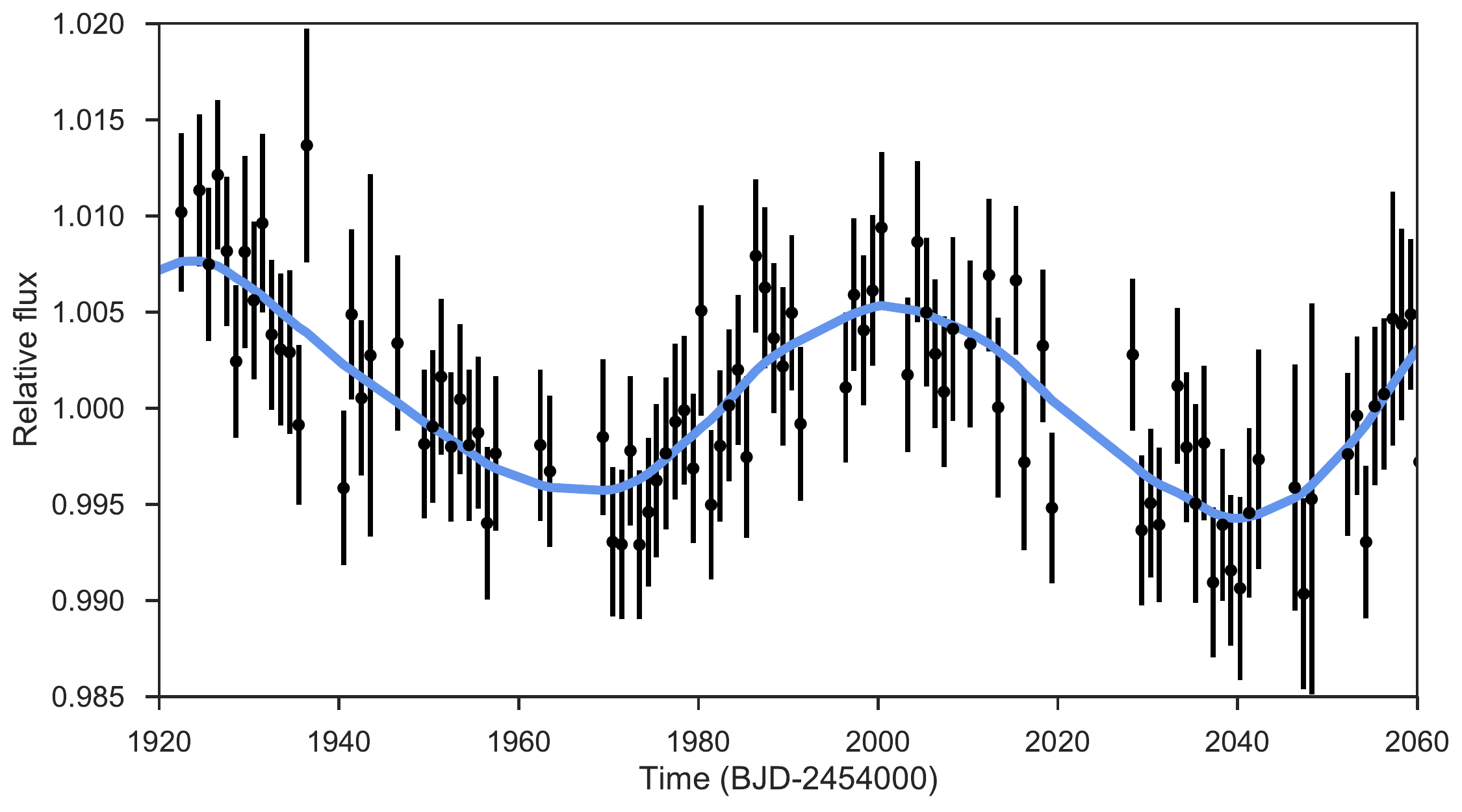}
    \caption{Close-up of the GP fit to all the photometric datasets used to estimate the stellar rotation period of the star. Black points show the WASP data, where a QP modulation can be clearly seen. Our best-fit GP fit (blue) reveals a rotational period of $P_\textnormal{rot} = 77.8^{+2.1}_{-2.0}$\,d.} 
    \label{fig:wasp-fit}
\end{figure}

Then, we use a more sophisticated model to determine precisely the empirical rotational period of the star by fitting the full photometric dataset described in Sect.~\ref{subsec:phot} (i.e., the ASAS, NSVS, ASAS-SN --- with observations both in $g$ and $V$ bands --- and WASP datasets) with a quasi-periodic (QP) Gaussian process (GP). In particular, we use the GP kernel introduced in \cite{celerite} of the form 
\begin{equation*}
k_{i,j}(\tau) = \frac{B}{2+C}e^{-\tau/L}\left[\cos \left(\frac{2\pi \tau}{P_\textnormal{rot}}\right) + (1+C)\right] \quad ,
\end{equation*}
where $\tau = |t_{i} - t_{j}|$ is the time-lag, $B$ and $C$ define the amplitude of the GP, $L$ is a timescale for the amplitude-modulation of the GP, and $P_\textnormal{rot}$ is the rotational period of the QP modulations. For the fit, we consider that each of the five datasets can have different values of $B$ and $C$ in order to account for the possibility that different bands could have different GP amplitudes, while the timescale of the modulation as well as the rotational period is left as a common parameter between the datasets. In addition, we fit for a flux offset between the photometric datasets, as well as for extra jitter terms added in quadrature to the diagonal of the resulting covariance matrix implied by this QP GP. We consider wide priors for $B$, $C$ (log-uniform between $10^{-5}$ ppm and $10^5$\,ppm), $L$ (log-uniform between $10^{-5}$ and $10^5$\,d), rotation period (uniform between 0 and 100\,d), flux offsets (Gaussian centered on 0 and standard deviation of $10^5$\,ppm), and jitters (log-uniform between 1 and $10^5$\,ppm). The fit is performed using \texttt{juliet} \citep[][ see next section for a full description of the algorithm]{juliet} and a close-up of the resulting fit is presented in \autoref{fig:wasp-fit} for illustration on how large the QP variations are in the WASP photometry, where the flux variability can be readily seen by eye. 

The resulting rotational period from this analysis is of $P_\textnormal{rot} = 77.8^{+2.1}_{-2.0}$\,d, consistent with the expectation from the small value of $\log R'_{HK}$.

\section{Analysis and results} \label{sec:fit}

\begin{figure*}
    \centering
    \includegraphics[width=\hsize]{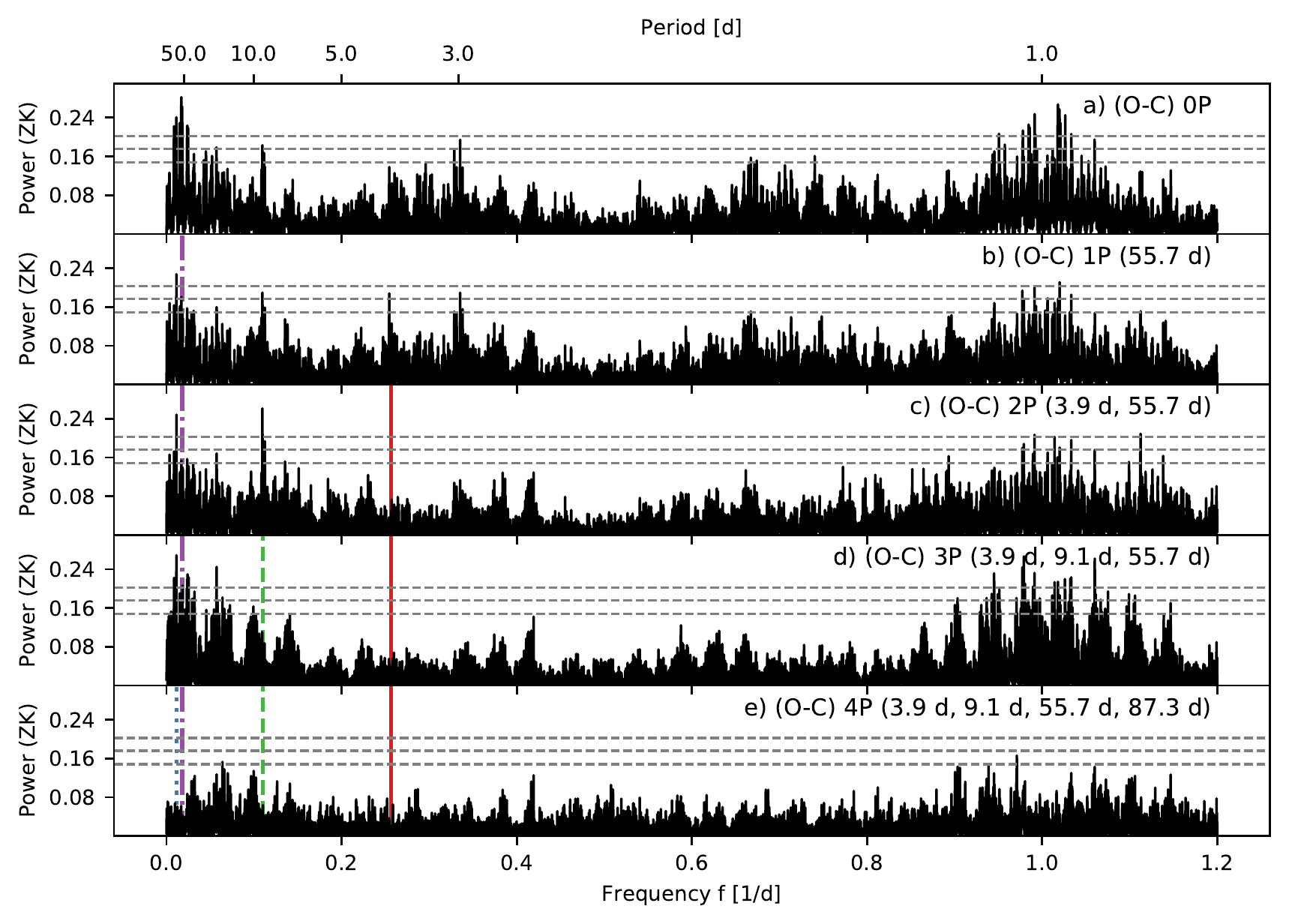}
    \caption{Generalized Lomb-Scargle periodograms of the residual RVs after subtraction of different models. Panel a: No signal subtracted, only instrumental offsets and jitter fitted. Panel b: Periodogram of the RV residuals after the subtraction of one sinusoidal signal with $P = 55.7\,\mathrm{d}$ (vertical purple dashed-dotted line). Panel c: Periodogram of the RV residuals after the simultaneous modeling of two signals with periods at 3.93\,d (red solid line) and $55.7\,\mathrm{d}$. Panel d: Periodogram of the RV residuals after the simultaneous modeling of three periodic signals with $P = 55.7\,$d, $P = 3.93\,$d, and $P = 9.1\,$d (green dashed line). Panel e: Periodogram of the RV residuals after the simultaneous modeling of four periodic signals with $P = 55.7$\,d, $P = 3.93$\,d, $P = 9.1\,$d, and $P \sim 87$\,d (blue dotted line).  The gray dashed lines indicate from bottom to top the analytic 10\%, 1\%, and 0.1\% FAP levels, respectively.}
    \label{fig:gls_rv}
\end{figure*}

\subsection{Period analysis of the RV data} \label{subsec:gls}

We performed a signal search in the RV data using generalized Lomb-Scargle (GLS) periodograms \citep{GLS}. Figure~\ref{fig:gls_rv} presents a series of GLS periodograms of the residual RVs after subtracting an increasing number of periodic signals. For each panel, we computed the theoretical false alarm probability (FAP) as described in \citet{GLS}, and show the 10\%, 1\%, and 0.1\% levels. After subtracting a model that fits only the instrumental offsets $\mu_{\rm instr}$ and jitters $\sigma_{{\rm instr}}$ (Fig.~\ref{fig:gls_rv}a), we find that the periodogram is dominated by a periodic signal at $P \sim 56\,\mathrm{d}$ and its aliases around periods of one day due to the sampling of the data.

After fitting a sinusoid to this signal, a GLS periodogram of the residuals shows many signals with $\mathrm{FAP} < 1\%$. One of those signals is at 3.93\,d, corresponding to the transiting planet detected in the {\it TESS} data. In this case, however, we want to know what is the probability that noise can produce a peak higher than what is seen exactly at the known frequency of the transiting planet, the spectral FAP. Following \citet{GLS}, we use a bootstrapping randomization method over a narrow frequency range centered on the planet orbital frequency to determine it. The analysis yields spectral $\mathrm{FAP} =  0.00075$. We thus estimate a $\mathrm{FAP} \sim 0.08\%$ for the 3.93 d signal.  

The residuals after the modeling of the 56\,d and 3.93\,d signals support a further periodicity of $P=9.1\,\mathrm{d}$ with a $\mathrm{FAP} < 0.1\%$ (Fig.~\ref{fig:gls_rv}c). This signal is persistent throughout the complete analysis and therefore cannot be explained by any of the other two known sources. Including this periodicity in the model as an extra sinusoid (Fig.~\ref{fig:gls_rv}d), the GLS of the residuals reveals a single relevant periodicity at 87\,d. When including a fourth sinusoid in the analysis at 87\,d, different peaks with $\mathrm{FAP} \approx 10\%$ populate the 1\,d region. We discuss in depth the nature of the four signals detected in the GLS in the next section, using more sophisticated models to fit the numerous periodicities in the RV dataset.

\subsection{Modeling results} \label{subsec:fit}

We used the recently published algorithm \texttt{juliet} \citep{juliet} to model jointly the photometric and Doppler data. The algorithm is built on many publicly available tools for the modeling of transits \citep[\texttt{batman},][]{batman}, RVs \citep[\texttt{radvel},][]{radvel}, and GP (\texttt{george}, \citealt{Ambikasaran2015ITPAM..38..252A}; \texttt{celerite}, \citealt{celerite}). In order to compare different models, \texttt{juliet} efficiently computes the Bayesian model log evidence ($\ln Z$) using either MultiNest \citep{multinest} via the \texttt{PyMultiNest} package \citep{pymultinest} or the \texttt{dynesty} package \citep{dynesty}. Nested sampling algorithms sample directly from the given priors instead of starting off with an initial parameter vector around a likelihood maximum found via optimization techniques, as done in common sampling methods. The trade-off between its versatility and completeness in the parameter space search is the computation time. For this reason, our prior choices have been selected to be the ideal balance between being informed, yet wide enough to fully acquire the posterior distribution map. We consider a model to be moderately favored over another if the difference in its Bayesian log evidence is greater than two, and strongly favored if it is greater than five \citep{2008ConPh..49...71T}. If $\Delta \ln Z \lesssim 2$, then the models are indistinguishable so the simpler model with less degrees of freedom would be chosen.

\subsubsection{Photometry only} \label{subsubsec:photonly}

In order to constrain the orbital period and time of transit center, we performed an analysis with \texttt{juliet} using only the \textit{TESS} photometry. We chose the priors in the orbital parameters from the \textit{TESS} DVR and our independent optimal BLS analysis. We adopted a few parametrization modifications when dealing with the transit photometry. Namely, we assigned a quadratic limb-darkening law for \textit{TESS}, as shown to be appropriate for space-based missions \citep{espinoza_jordan_limb:2015}, which then was parametrized with the uniform sampling scheme ($q_1, q_2$), introduced by \citet{Kipping13}. Additionally, rather than fitting directly for the planet-to-star radius ratio ($p=R_p/R_*$) and the impact parameter of the orbit ($b$), we instead used the parametrization introduced in \cite{Espinoza18} and fit for the parameters $r_1$ and $r_2$ to guarantee full exploration of physically plausible values in the ($p, b$) plane. Lastly, we applied the classical parametrization of ($e$, $\omega$) into ($\mathcal{S}_1 = \sqrt{e}\sin\omega$, $\mathcal{S}_2 = \sqrt{e}\cos\omega$), always ensuring that $e = \mathcal{S}_1^2 + \mathcal{S}_2^2 \leq 1$. We fixed the \textit{TESS} dilution factor to one based on our analysis from Sects.~\ref{subsec:contamination} and \ref{subsec:fastcam}, but accounted for any residual time-correlated noise in the light curve with an exponential GP kernel of the form $k_{i,j} = \sigma^2_\mathrm{GP,TESS} \exp\left(- |t_i - t_j|/T_\mathrm{GP,TESS}\right)$, where $T_\mathrm{GP,TESS}$ is a characteristic timescale and $\sigma_\mathrm{GP,TESS}$ is the amplitude of this GP modulation. Furthermore, we added in quadrature a jitter term $\sigma_{\mathrm{TESS}}$ to the {\it TESS} photometric uncertainties, which might be underestimated due to additional systematics in the space-based photometry. The details of the priors and the description for each parameter are presented in Table~\ref{tab:priors} of the Appendix.

The results from the photometry-only analysis with \texttt{juliet} are completely consistent with those provided by the \textit{TESS} DVR and our independent transit search, but with improved precision in the transit parameters after accounting for extra systematics with the jitter term and the GP. We also searched for an additional transiting planet in the system by modeling a two-planet fit where we use the same priors in Table~\ref{tab:priors} for the first planet, and then allow the period and time of transit center to vary for the second. The transiting model for the hypothetical second planet is totally flat and we find no strong evidence ($\Delta \ln Z = \ln Z_{\rm 1pl} - \ln Z_{\rm 2pl}  = 5.16$) for any additional transiting planets in the light curve, in agreement with our findings in \autoref{subsec:transit}.

\subsubsection{RV only} \label{subsubsec:rvonly}

\begin{table}
    \centering
    \caption{Model comparison of RV-only fits with \texttt{juliet}. The prior label $\mathcal{N}$ represents a normal distribution. The final model used for the joint fit is marked in boldface (see Sect.~\ref{subsubsec:rvonly} for details about the selection of the final model). }  \label{tab:models}
    \begin{tabular}{llcc}
        \hline
        \hline
        \noalign{\smallskip}
        Model & Prior $P_{\rm planet}$ & GP kernel  & $\Delta\ln Z$    \\
        \noalign{\smallskip}
        \hline
        \noalign{\smallskip}
1pl     & $\mathcal{N}_\mathrm{b}(55.7,0.5^2)$        & \dots     & 19.93      \\[0.1cm]
2pl     & $\mathcal{N}_\mathrm{b}(3.931,0.001^2)$   & \dots     & 15.51      \\
        & $\mathcal{N}_\mathrm{c}(55.7,0.5^2)$        &           &          \\[0.1cm]
3pl     & $\mathcal{N}_\mathrm{b}(3.931,0.001^2)$   & \dots     & ~9.03      \\
        & $\mathcal{N}_\mathrm{c}(9.1,0.1^2)$         &           &          \\
        & $\mathcal{N}_\mathrm{d}(55.7,0.5^2)$        &           &          \\[0.1cm]
4pl     & $\mathcal{N}_\mathrm{b}(3.931,0.001^2)$   & \dots     & -2.15      \\
        & $\mathcal{N}_\mathrm{c}(9.1,0.1^2)$         &           &          \\
        & $\mathcal{N}_\mathrm{d}(55.7,0.5^2)$        &           &          \\
        & $\mathcal{N}_\mathrm{e}(87.3,0.5^2)$        &           &          \\
        \noalign{\smallskip}
        \noalign{\smallskip}
1pl+GPexp   & $\mathcal{N}_\mathrm{b}(3.931,0.001^2)$   & Exp\tablefootmark{a}     & 19.51      \\[0.1cm]
2pl+GPexp   & $\mathcal{N}_\mathrm{b}(3.931,0.001^2)$   & Exp\tablefootmark{a}     & ~9.29     \\
            & $\mathcal{N}_\mathrm{c}(9.1,0.1^2)$         &                               &         \\[0.1cm]
{\bf 3pl+GPexp} & $\mathcal{N}_\mathrm{b}(3.931,0.001^2)$   & Exp\tablefootmark{a}     & {\bf ~0.00 }     \\
            & $\mathcal{N}_\mathrm{c}(9.1,0.1^2)$         &                               &         \\
            & $\mathcal{N}_\mathrm{d}(55.7,0.5^2)$        &                               &         \\
4pl+GPexp   & $\mathcal{N}_\mathrm{b}(3.931,0.001^2)$   & Exp\tablefootmark{a}     & -0.95     \\
            & $\mathcal{N}_\mathrm{c}(9.1,0.1^2)$         &                               &         \\
            & $\mathcal{N}_\mathrm{d}(55.7,0.5^2)$        &                               &         \\
            & $\mathcal{N}_\mathrm{e}(87.3,0.5^2)$        &           &          \\
        \noalign{\smallskip}
        \noalign{\smallskip}
1pl+GPess  & $\mathcal{N}_\mathrm{b}(3.931,0.001^2)$   & ExpSinSq\tablefootmark{b}     & 18.06      \\[0.1cm]
2pl+GPess  & $\mathcal{N}_\mathrm{b}(3.931,0.001^2)$   & ExpSinSq\tablefootmark{b}     & ~6.32      \\
            & $\mathcal{N}_\mathrm{c}(9.1,0.1^2)$         &                               &         \\[0.1cm]
3pl+GPess  & $\mathcal{N}_\mathrm{b}(3.931,0.001^2)$   & ExpSinSq\tablefootmark{b}     & -1.59     \\
            & $\mathcal{N}_\mathrm{c}(9.1,0.1^2)$         &                               &         \\
            & $\mathcal{N}_\mathrm{d}(55.7,0.5^2)$        &                               &         \\
4pl+GPess  & $\mathcal{N}_\mathrm{b}(3.931,0.001^2)$   & ExpSinSq\tablefootmark{b}     & ~3.11     \\
            & $\mathcal{N}_\mathrm{c}(9.1,0.1^2)$         &                               &         \\
            & $\mathcal{N}_\mathrm{d}(55.7,0.5^2)$        &                               &         \\
            & $\mathcal{N}_\mathrm{e}(87.3,0.5^2)$        &           &          \\
        \noalign{\smallskip}
        \hline
    \end{tabular}
    \tablefoot{
        \tablefoottext{a}{Simple exponential kernel of the form $k_{i,j} = \sigma^2_\mathrm{GP,RV} \exp\left(-|t_i - t_j|/T_\mathrm{GP,RV}\right)$}.
        \tablefoottext{b}{Exponential-sine-squared kernel of the form $k_{i,j} = \sigma^2_\mathrm{GP,RV} \exp\left(- \alpha_\mathrm{GP,RV} (t_i - t_j)^2 - \Gamma_\mathrm{GP,RV} \sin^2 \left[\frac{\pi |t_i - t_j|}{P_{\rm rot;GP,RV}}\right]\right)$ with a uniform prior in $P_{\rm rot;GP,RV}$ ranging from 30 to 100\,d.} 
    }
\end{table}

In \autoref{subsec:gls}, we have shown that several signals are present in the RV data. We tested several models using \texttt{juliet} on the RV dataset independently to understand the nature of those signals and their significance when doing a simultaneous multi-planetary fit. We discuss three sets of models, each exploring possible system architectures covering the four interesting periodicities from \autoref{subsec:gls} of 3.93\,d, 9.1\,d, 55.7\,d, and 87.3\,d. The details regarding the priors and Bayesian log evidence of all the runs are listed in \autoref{tab:models}. We included an instrumental jitter term for each of the six individual RV datasets and assumed circular orbits. We also considered eccentric orbits but found the circular model fits to have comparable log evidence and be computationally less expensive.

The first set of models (1pl, 2pl, 3pl, 4pl) treats the signals found in the periodogram analysis as Keplerian circular orbits. The preferred model is clearly the four-planet one, with a minimum $\Delta\ln Z > 11$ with respect to the others. This model is also the one with the highest evidence in our analysis. The three signals at 3.93\,d, 9.1\,d, and 55.7\,d have eccentricities compatible with zero. However, the derived eccentricity for the 87.3\,d signal is substantially high ($e \sim 0.4$). We notice that the RV phase-folded curve to the 87.3\,d signal is not homogeneously sampled, with very few RV points covering both quadratures, which could explain the relatively high eccentric behavior derived for the 87.3\,d signal.  

To test the planetary nature of the signals, we also tried more complex models using GP regression to account for correlated noise. The explicit mathematical form of the GP kernels can be found in the notes to Table~\ref{tab:models}. We first employed a quasi-periodic exponential-sine-squared kernel (GPess) using a wide prior for the period term. In doing this, we can evaluate the preference for Keplerian signals over correlated periodic noise in the data, especially focusing on modeling the dubious 87.3\,d periodicity. Even though the posterior distribution does not show any interesting signals, a periodogram of the GP component of the 3pl+GPess model -- obtained by substracting the median Keplerian model to our full median 3pl+GPess model -- reveals that the 87.3\,d periodicity is the main component. However, if we consider the 87.3\,d as a Keplerian along with the other three periodicities and a GPess kernel to account for the residual noise seen in Fig.~\ref{fig:gls_rv}e, it yields worse results in terms of model log evidence. Besides, a simpler exponential kernel (GPexp) for the GP with a three-Keplerian model gives comparable results to the 3pl+GPess model. This indicates that the exponential kernel is sufficient in accounting for the stochastic behavior of the data. Additionally, we tried a GPess kernel with a normal prior in $P_{\rm rot;GP,RV}$ centered at the 77.8\,d rotational period of the star derived in Sect.~\ref{subsec:rotation}. In that case, the results are worse or equivalent to the GPexp kernel case, meaning that the GP model did not catch any clear periodicity. This suggests also that the stellar spots do not imprint any modulation in the RVs, which is in line with the absence of peaks around 78\,d in the periodograms of Fig.~\ref{fig:gls_rv}.

Given the log evidence of the different models in Table~\ref{tab:models}, 4pl, 3pl+GPexp, and 3pl+GPess are statistically equivalent. Although 4pl is weakly favored in terms of $\ln Z$, the fact that the derived orbit of the 87.3\,d signal is much more eccentric than the others and the phase is not well covered by our measurements withdraw us from firmly claiming the signal as of planetary nature. Further observations of GJ~357 will help shed light on the true nature of this signal and further potential candidates. Therefore, for the final joint fit we consider the model with three Keplerians and an exponential GP to be the simplest model that best explains the current data present.

\subsubsection{Joint fit} \label{subsubsec:joint}

\begin{figure*}
    \centering
    \includegraphics[width=\hsize]{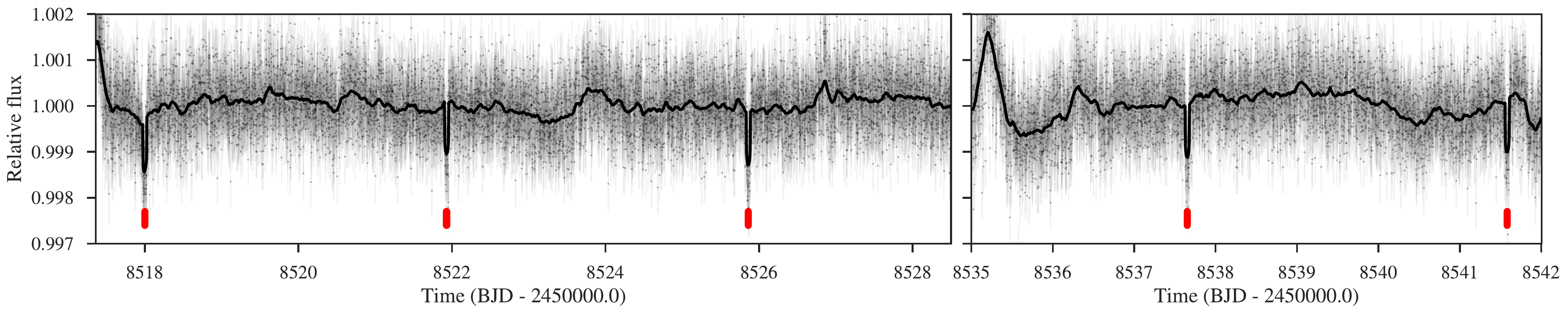} \\
    \begin{minipage}{.4\textwidth}
        \centering
        \includegraphics[width=1\linewidth]{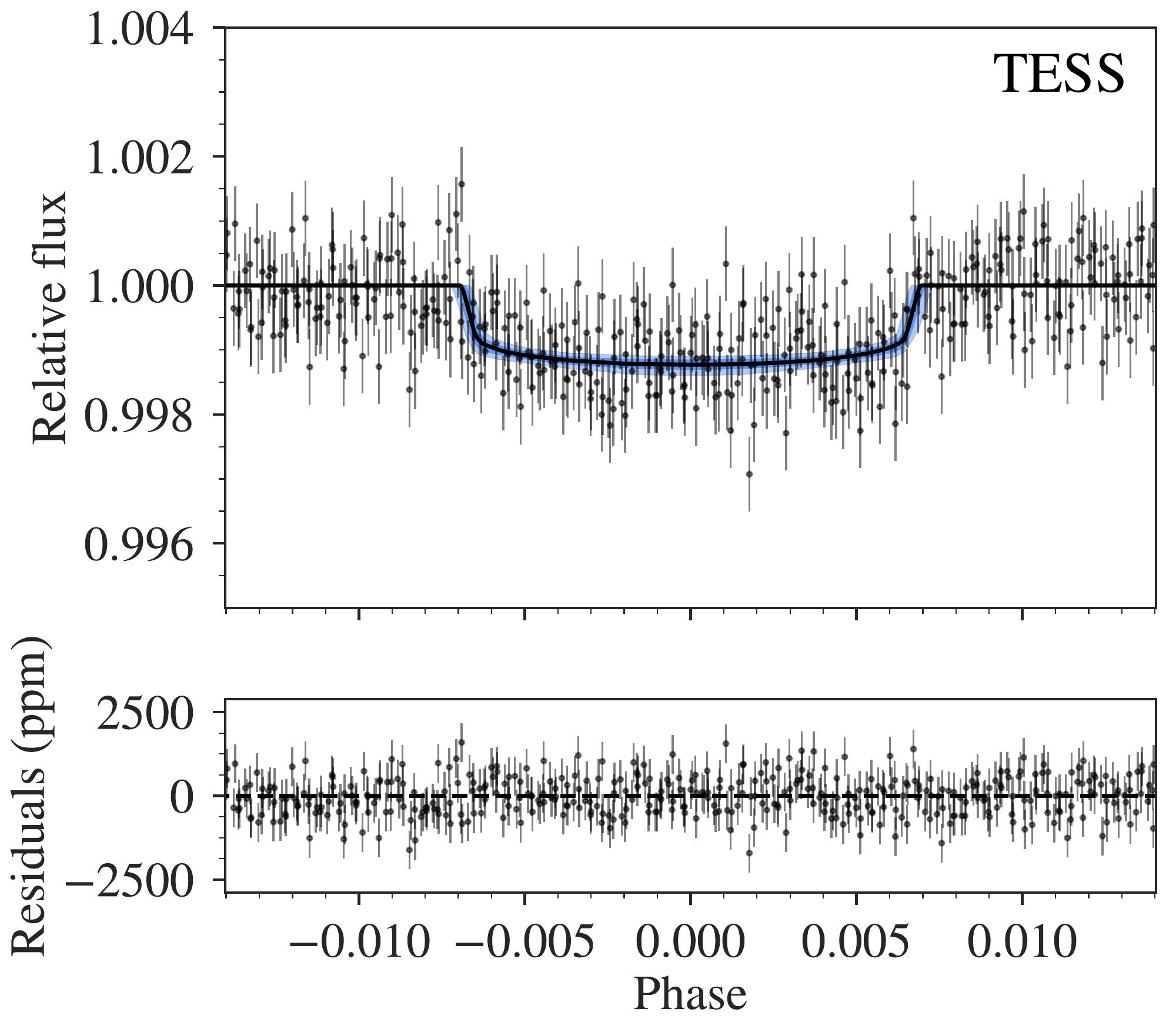}
    \end{minipage}
    \begin{minipage}{.4\textwidth}
        \centering
        \includegraphics[width=1\linewidth]{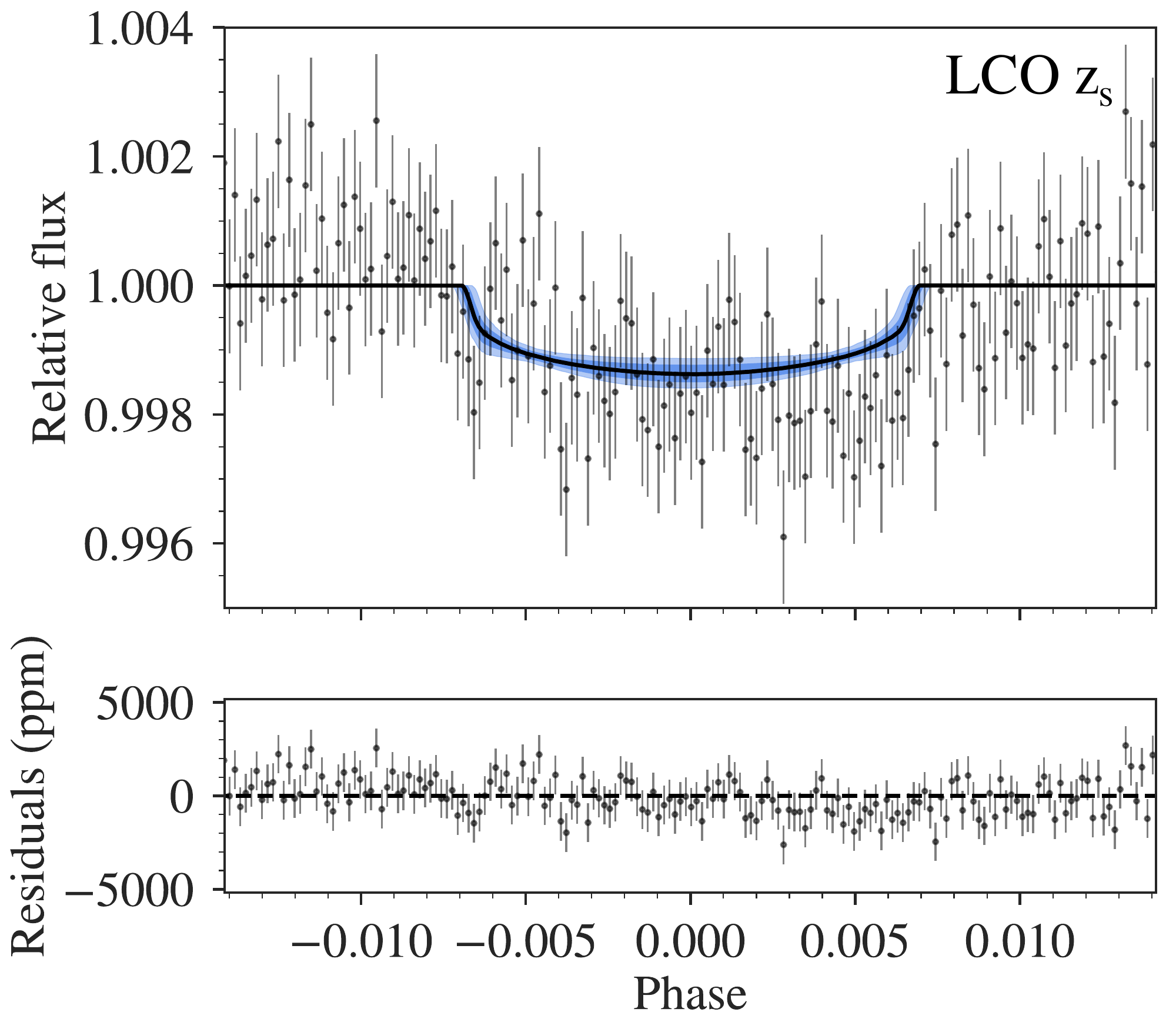}
    \end{minipage}
    \includegraphics[width=\hsize]{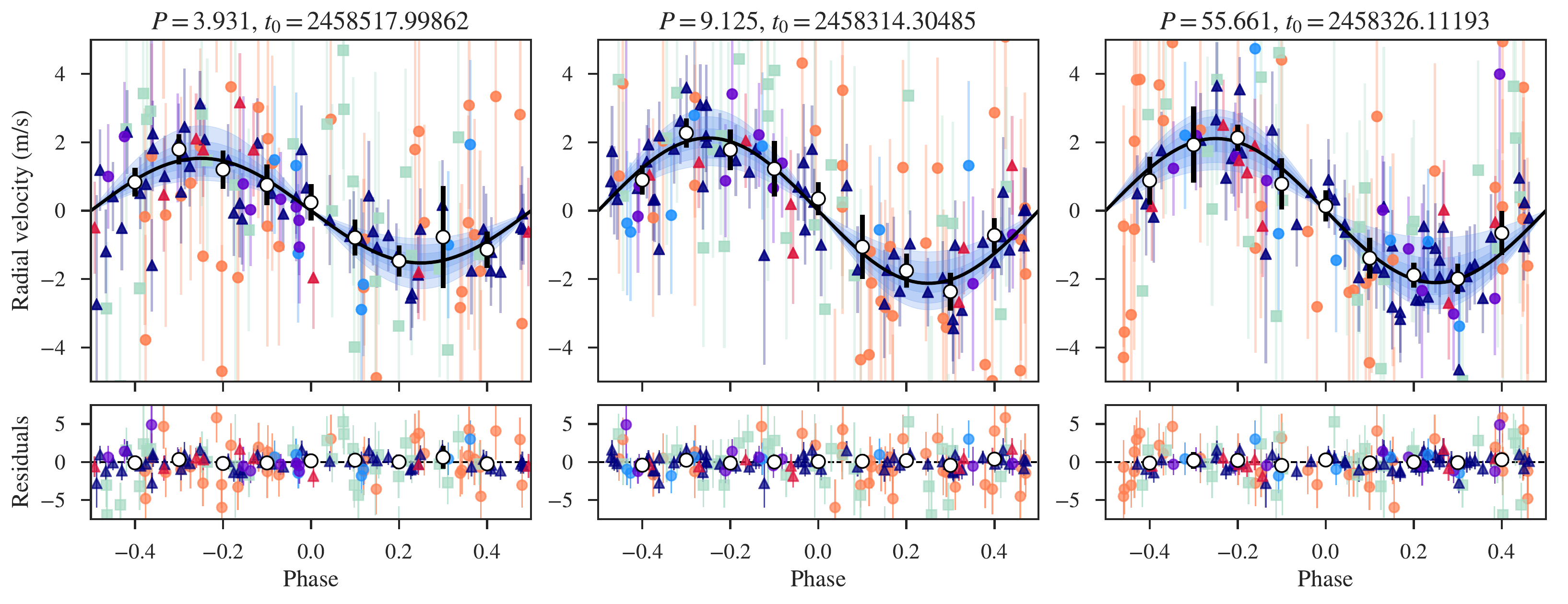}
    \caption{Results from the joint fit of the best model 3pl+GPexp. Top panel: {\it TESS} photometry time series (gray points with error bars) along with the best-fit model (solid black line) from our joint modeling. This best-fit model includes an exponential GP used to account for the evident trends as well as for a transit model. Individual transits of GJ~357~b are indicated with red ticks. Middle panel: {\it TESS} photometry (left) and LCO photometry (right) phase-folded to the 3.93\,d period of GJ~357~b along with best-fit transit model from the joint fit. The GP fitted to the photometry has been removed. Bottom panel: RVs phase-folded to the period of the three confirmed planets (GJ~357~b, left; GJ~357~c, center; GJ~357~d, right). RV data come from HIRES (orange circles), UVES (light green squares), HARPS (navy blue triangles), PFSpre (light blue circles), PFSpost (red triangles), and CARMENES (purple circles). The GP fitted to the RV dataset has been removed. White circles show binned datapoints in phase for visualisation. The error bars of both photometry and RV data include their corresponding jitter.} 
    \label{fig:jointfit}
\end{figure*}

\begin{table}
    \centering
    \caption{Posterior parameters of the final joint fit obtained for GJ~357~b,~c, and~d, using \texttt{juliet}. Priors and descriptions for each parameter can be found in Table~\ref{tab:priors}.}
    \label{tab:posteriors}
    \begin{tabular}{l@{\hspace{-3mm}}c@{\hspace{3mm}}c@{\hspace{3mm}}c@{\hspace{3mm}}} 
        \hline
        \hline
        \noalign{\smallskip}
        Parameter$^{(a)}$ & GJ~357~b & GJ~357~c & GJ~357~d \\
        \noalign{\smallskip}
        \hline
        \noalign{\smallskip}
        \multicolumn{4}{c}{\it Stellar parameters} \\[0.1cm]
        \noalign{\smallskip}
        $\rho_\star$ ($\mathrm{kg\,m\,^{-3}}$)       & \multicolumn{3}{c}{$13600.0^{+1400}_{-1600}$} \\[0.1 cm]
        \noalign{\smallskip}
        \multicolumn{4}{c}{\it Planet parameters} \\[0.1cm]
        \noalign{\smallskip}
        $P$ (d)                                & $3.93072^{+0.00008}_{-0.00006}$ & $9.1247^{+0.0011}_{-0.0010}$ & $55.661^{+0.055}_{-0.055}$ \\[0.1 cm]
        $t_0$\tablefootmark{(b)}                  & $8517.99862^{+0.00039}_{-0.00038}$ & $8314.30^{+0.42}_{-0.38}$ & $8326.1^{+3.9}_{-3.8}$ \\[0.1 cm]
        $r_1$                                & $0.56^{+0.07}_{-0.09}$ & \dots & \dots \\[0.1 cm]
        $r_2$                                & $0.0331^{+0.0009}_{-0.0009}$ & \dots & \dots \\[0.1 cm]
        $K$ ($\mathrm{m\,s^{-1}}$)           & $1.52^{+0.25}_{-0.25}$ & $2.13^{+0.28}_{-0.28}$ & $2.09^{+0.34}_{-0.35}$ \\[0.1 cm]
        \noalign{\smallskip}
        \multicolumn{4}{c}{\it Photometry parameters} \\[0.1cm]
        \noalign{\smallskip}
        $M_{\textnormal{TESS}}$ (ppm)            & \multicolumn{3}{c}{$-47^{+65}_{-69}$} \\[0.1 cm]
        $M_{\textnormal{LCO}}$ (ppm)            & \multicolumn{3}{c}{$173^{+65}_{-64}$} \\[0.1 cm]
        $\sigma_{\textnormal{TESS}}$ (ppm)     & \multicolumn{3}{c}{$127^{+15}_{-16}$} \\[0.1 cm] 
        $\sigma_{\textnormal{LCO}}$ (ppm)     & \multicolumn{3}{c}{$928^{+50}_{-48}$} \\[0.1 cm] 
        $q_{1,\textnormal{TESS}}$                & \multicolumn{3}{c}{$0.20^{+0.21}_{-0.13}$} \\[0.1 cm]
        $q_{2,\textnormal{TESS}}$                & \multicolumn{3}{c}{$0.32^{+0.34}_{-0.21}$} \\[0.1 cm]
        $q_{1,\textnormal{LCO}}$                & \multicolumn{3}{c}{$0.72^{+0.18}_{-0.28}$} \\[0.1 cm]
        \noalign{\smallskip}
        \multicolumn{4}{c}{\it RV parameters} \\[0.1cm]
        \noalign{\smallskip}
        $\mu_{\textnormal{HIRES}}$ ($\mathrm{m\,s^{-1}}$)            & \multicolumn{3}{c}{$0.96^{+0.60}_{-0.62}$} \\[0.1 cm]
        $\sigma_{\textnormal{HIRES}}$ ($\mathrm{m\,s^{-1}}$)       & \multicolumn{3}{c}{$1.99^{+0.78}_{-0.85}$} \\[0.1 cm]
        $\mu_{\textnormal{UVES}}$ ($\mathrm{m\,s^{-1}}$)             & \multicolumn{3}{c}{$1.06^{+0.64}_{-0.66}$} \\[0.1 cm]
        $\sigma_{\textnormal{UVES}}$ ($\mathrm{m\,s^{-1}}$)        & \multicolumn{3}{c}{$1.68^{+0.89}_{-0.90}$} \\[0.1cm]
        $\mu_{\textnormal{HARPS}}$ ($\mathrm{m\,s^{-1}}$)            & \multicolumn{3}{c}{$-5.15^{+0.41}_{-0.40}$} \\[0.1 cm]
        $\sigma_{\textnormal{HARPS}}$ ($\mathrm{m\,s^{-1}}$)       & \multicolumn{3}{c}{$0.61^{+0.41}_{-0.37}$} \\[0.1 cm]
        $\mu_{\textnormal{PFSpre}}$ ($\mathrm{m\,s^{-1}}$)            & \multicolumn{3}{c}{$-2.31^{+1.17}_{-1.19}$} \\[0.1 cm]
        $\sigma_{\textnormal{PFSpre}}$ ($\mathrm{m\,s^{-1}}$)       & \multicolumn{3}{c}{$1.71^{+1.39}_{-1.04}$} \\[0.1 cm]
        $\mu_{\textnormal{PFSpost}}$ ($\mathrm{m\,s^{-1}}$)            & \multicolumn{3}{c}{$-0.97^{+0.99}_{-0.99}$} \\[0.1 cm]
        $\sigma_{\textnormal{PFSpost}}$ ($\mathrm{m\,s^{-1}}$)       & \multicolumn{3}{c}{$1.20^{+0.99}_{-0.74}$} \\[0.1 cm]
        $\mu_{\textnormal{CARM}}$ ($\mathrm{m\,s^{-1}}$)         & \multicolumn{3}{c}{$-1.61^{+0.93}_{-0.92}$} \\[0.1 cm]
        $\sigma_{\textnormal{CARM}}$ ($\mathrm{m\,s^{-1}}$)    & \multicolumn{3}{c}{$0.98^{+0.99}_{-0.64}$} \\[0.1cm]
        \noalign{\smallskip}
        \multicolumn{4}{c}{\it GP hyperparameters} \\
        \noalign{\smallskip}
        $\sigma_\mathrm{GP,TESS}$ (ppm)           & \multicolumn{3}{c}{$0.10^{+0.04}_{-0.02}$} \\[0.1 cm]
        $T_\mathrm{GP,TESS}$ (d)                  & \multicolumn{3}{c}{$0.38^{+0.17}_{-0.09}$} \\[0.1 cm]
        $\sigma_\mathrm{GP,RV}$ ($\mathrm{m\,s^{-1}}$)              & \multicolumn{3}{c}{$2.66^{+1.02}_{-0.76}$} \\[0.1 cm]
        $T_\mathrm{GP,RV}$ (d)                    & \multicolumn{3}{c}{$0.12^{+0.12}_{-0.06}$} \\[0.1 cm]

        \noalign{\smallskip}
        \hline
    \end{tabular}
    \tablefoot{
        \tablefoottext{a}{Error bars denote the $68\%$ posterior credibility intervals.}
        \tablefoottext{b}{Units are BJD - 2450000.}
    }
\end{table}

\begin{table}
    \centering
    \caption{Derived planetary parameters obtained for GJ~357~b, c, and d using the posterior values from Table~\ref{tab:posteriors}.}
    \label{tab:derivedparams}
    \begin{tabular}{lccc} 
        \hline
        \hline
        \noalign{\smallskip}
        Parameter\tablefootmark{(a)} & GJ~357~b & GJ~357~c & GJ~357~d \\
        \noalign{\smallskip}
        \hline
        \noalign{\smallskip}
        \multicolumn{4}{c}{\it Derived transit parameters} \\[0.1cm]
        \noalign{\smallskip}
        $p = R_{\rm p}/R_\star$            & $0.0331^{+0.0009}_{-0.0009}$ & \dots & \dots \\[0.1 cm]
        $b = (a/R_\star)\cos i_{\rm p}$    & $0.34^{+0.10}_{-0.14}$       & \dots & \dots \\[0.1 cm]
        $a/R_\star$                  & $22.31^{+0.76}_{-0.90}$      & \dots & \dots \\[0.1 cm]
        $i_p$ (deg)              & $89.12^{+0.37}_{-0.31}$      & \dots & \dots \\[0.1 cm]
        $u_1$\tablefootmark{(b)}             & $0.27^{+0.24}_{-0.17}$       & \dots & \dots \\[0.1 cm]
        $u_2$\tablefootmark{(b)}             & $0.14^{+0.29}_{-0.24}$       & \dots & \dots \\[0.1 cm]
        $t_T$ (h)                & $1.53^{+0.12}_{-0.11}$       & \dots & \dots \\[0.1 cm]
        \noalign{\smallskip}
        \multicolumn{4}{c}{\it Derived physical parameters} \\[0.1cm]
        \noalign{\smallskip}
        $M_{\rm p}$ ($M_\oplus$)\tablefootmark{(c)}    & $1.84^{+0.31}_{-0.31}$   & $> 3.40^{+0.46}_{-0.46}$ & $> 6.1^{+1.0}_{-1.0}$ \\[0.1 cm]
        $R_{\rm p}$ ($R_\oplus$)        & $1.217^{+0.084}_{-0.083}$   & \dots & \dots \\[0.1 cm]
        $\rho_{\rm p}$ (g cm$^{-3}$)             & $5.6^{+1.7}_{-1.3}$      & \dots & \dots \\[0.1 cm]
        $g_{\rm p}$ (m s$^{-2}$)                 & $12.1^{+2.9}_{-2.5}$     & \dots & \dots \\[0.1 cm]
        $a_{\rm p}$ (au)                         & $0.035^{+0.002}_{-0.002}$ & $0.061^{+0.004}_{-0.004}$ & $0.204^{+0.015}_{-0.015}$ \\[0.1 cm]
        $T_\textnormal{eq}$ (K)\tablefootmark{(d)}          & $525^{+11}_{-9}$   & $401.2^{+10.8}_{-10.7}$ & $219.6^{+5.9}_{-5.9}$ \\[0.1 cm]
        $S$ ($S_\oplus$)            & $12.6^{+1.1}_{-0.8}$  & $4.45^{+0.14}_{-0.14}$ & $0.38^{+0.01}_{-0.01}$ \\[0.1 cm]
        \noalign{\smallskip}
        \hline
    \end{tabular}
    \tablefoot{
      \tablefoottext{a}{Error bars denote the $68\%$ posterior credibility intervals.}
      \tablefoottext{b}{Derived only from the {\it TESS} light curve.}
      \tablefoottext{c}{The masses for GJ~357~c and GJ~357~d are a lower limit ($M_{\rm p} \sin i$) since they are detected in the RV data only.}
      \tablefoottext{d}{Equilibrium temperatures were calculated assuming zero Bond albedo.}
      }
\end{table}

To obtain the most precise parameters of the GJ~357 system, we performed a joint analysis of the {\it TESS} and LCO photometry and Doppler data using \texttt{juliet,} of the model 3pl+GPexp from our RV-only analysis in \autoref{subsubsec:rvonly}. In this way, we simultaneously constrain all the parameters for the transiting planet GJ~357~b, the planet candidates at 9.12\,d and 55.7\,d, and the correlated noise seen in the RV data with an exponential GP kernel. To optimize the computational time, we narrowed down our priors based on the analyses from the sections above, but we kept them wide enough to fully sample the posterior distribution of the quantities of interest. Our choice of the priors for each parameter in the joint analysis of the 3pl+GPexp model can be found in Table~\ref{tab:priors}.

The posterior distribution of the parameters of our best-fit joint model are presented in Table~\ref{tab:posteriors}. We also ran an eccentric version of the joint 3pl+GPexp model, but the circular fit model was strongly favored ($\Delta \ln Z  > 8$) and thus we only show the circular results in Table~\ref{tab:posteriors}. The corresponding modeling of the data based on these posteriors is shown in \autoref{fig:jointfit} and the derived physical parameters of the system are presented in \autoref{tab:derivedparams}. The RV time series is shown for completeness in the Appendix (Fig.~\ref{fig:rv_timeseries1}).

\section{Discussion} \label{sec:discussion}

The GJ~357 system consists of one transiting Earth-sized planet in a 3.93\,d orbit, namely GJ~357~b, with a radius of $R_{\rm b}=1.217\pm0.084\,R_\oplus$, a mass of $M_{\rm b}=1.84\pm0.31 M_\oplus$, and a density of $\rho_{\rm b}=5.6^{+1.7}_{-1.3}\,\mathrm{g\,cm^{-3}}$; and two additional planets, namely GJ~357~c, with a minimum mass of $M_{\rm c}=3.40\pm0.46\,M_\oplus$ in a 9.12\,d orbit, and GJ~357~d, with a minimum mass of $M_{\rm c}=6.1\pm1.0\,M_\oplus$ in a 55.7\,d orbit. The modulations from the GP model have an amplitude of $2.66\,\mathrm{m\,s^{-1}}$ and account for the short timescale of the stochastic variations and the dubious 87.3\,d signal. 

\subsection{Searching for transits of planets c and d }

Although in Sects.~\ref{subsec:transit} and \ref{subsubsec:photonly} we looked for additional transit features in the {\it TESS} light curve, but could not find any, we performed a last run with \texttt{juliet} to rule out the possibility that GJ~357~c transits. To do so, we took the period and time of transit center from Table~\ref{tab:posteriors} as priors and added $r_{1,{\rm c}}$ and $r_{2,{\rm c}}$ as free parameters assuming an eccentricity equal to zero. In agreement with previous analyses, the log evidence shows that the non-transiting model for GJ~357~c is slightly preferred ($\Delta \ln Z \sim 3$).

Therefore, we can conclude that GJ~357~c, although firmly detected in the RV dataset, does not transit. On the other hand, any possible transit of GJ~357~d would have been missed by {\it TESS} observations, according to the predicted transit epoch from the RVs fits. With the current data the uncertainty in the ephemeris is on the order of days, however, additional RVs could improve the precision on the period and phase determination enough to allow a transit search. The a priori transit probability is only 0.8\%, but this transit search is doable by {\it CHEOPS} \citep{CHEOPS} since the star will fall in its 50\,d observability window.

As in other planetary systems recently revealed by {\it TESS} \citep[e.g.,][]{Espinoza2019arXiv190307694E} or {\it Kepler}/{\it K2} \citep[e.g.,][]{Quinn2015ApJ...803...49Q, Buchhave2016AJ....152..160B, Otor2016AJ....152..165O, Christiansen2017AJ....154..122C, Gandolfi2017AJ....154..123G, Cloutier2017A&A...608A..35C}, non-transiting planets in these transit-detected systems expand our understanding of the system architecture. In those works, after the modeling of the transiting planet detected in the space-based photometry, further signals appear in the RV data that would not have been significant otherwise. This emphasizes the importance of long, systematic, and precise ground-based searches for planets around bright stars and the relevance of archival public data.

\subsection{System architecture}

\begin{figure}
    \centering
    \includegraphics[width=\hsize]{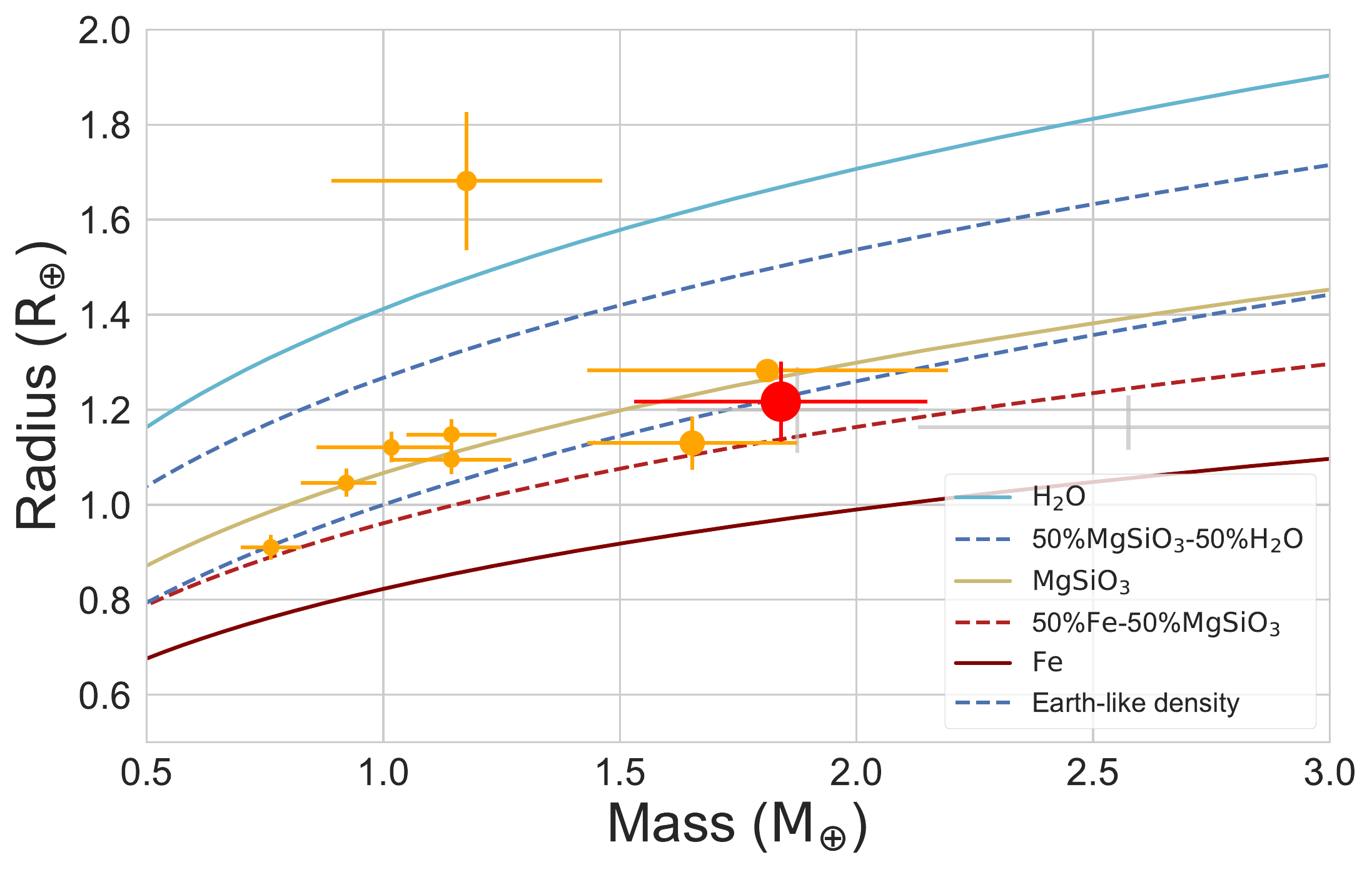}
    \caption{Mass-radius diagram for all known transiting planets with masses between 0.5--3\,$M_\oplus$ and radii 0.5--2\,$R_\oplus$ determined with a precision better than 30\%. GJ~357~b is shown in red. Planets orbiting around late-type stars ($T_{\rm eff} < 4000\,\mathrm{K}$) are shown in orange, otherwise gray. The size of the orange datapoints is inversely proportional to the magnitude of their host star in the $J$-band. Data are taken from the TEPCat database of well-characterized planets \citep{Southworth2011MNRAS.417.2166S}. Theoretical models for the planet's internal composition are taken from \citet{Zeng2016ApJ...819..127Z}.}
    \label{fig:massradius}
\end{figure}

We determine that the transiting planet GJ~357~b has a mass of $M_\mathrm{b} = 1.84\pm0.31\,M_\oplus$ and a radius of $R_\mathrm{b} = 1.217\pm0.084\,R_\oplus$, which corresponds to a bulk density of $\rho_\mathrm{b} = 5.6_{-1.3}^{+1.7}\,\mathrm{g~cm^{-3}}$. Figure~\ref{fig:massradius} shows masses and radii of all confirmed planets whose precision in both parameters is better than 30\%. GJ~357~b joins the very small group of Earth-sized and Earth-mass planets orbiting M dwarfs discussed in the introduction. The bulk density we measure for GJ~357~b overlaps with the 30\% Fe and 70\% MgSiO$_3$ mass-radius curve as calculated by \citet{Zeng2016ApJ...819..127Z}.

We derive a minimum mass for GJ~357~c of $M_\mathrm{c}\sin i_\mathrm{c} = 3.40\pm0.46\,M_\oplus$, which falls between the vaguely defined mass range for Earth- and super-Earth-like planets. Since we determine only a lower limit for $M_\mathrm{c}$ , it is possible that this planet falls into the super-Earth category joining the group of planets that make up the lower radius bump in the bimodal distribution of small planets found with {\it Kepler} \citep{2017AJ....154..109F, 2018MNRAS.479.4786V}. 

The derived minimum mass of GJ~357~d is $M_\mathrm{d}\sin i_\mathrm{d} = 6.1\pm1.0\,M_\oplus$. However, a prediction of its bulk composition is not straightforward since both super-Earth ($1<R<2\,R_\oplus$) and mini-Neptune ($2<R<4\,R_\oplus$) exoplanets encompass this range of masses. As an example of this dichotomy, Kepler-68~b \citep[][$R = 2.33 \pm 0.02\,R_\oplus$]{Kepler68} and Kepler-406~b \citep[][$R = 1.43 \pm 0.03\,R_\oplus$]{2014ApJS..210...20M} have similar masses of $M \sim 6.0\,M_\oplus$ and $M \sim 6.3\,M_\oplus$, respectively, but their compositions differ significantly (between rocky and gaseous for Kepler-68~b, and purely rocky for Kepler-406~b). 

We note that the planetary system of GJ~357 is quite similar to that of GJ~1132 \citep{GJ1132,Bonfils2018AA...618A.142B}. In both cases we find a similar bulk density of the inner planet of $\sim 6\,\mathrm{g~cm^{-3}}$ (see orange datapoint down left of GJ~357~b in Fig.~\ref{fig:massradius}) and a second non-transiting planet with an orbital period around 9\,d. Furthermore, the long-term trends found with GJ~1132 could indicate the presence of one or more outer, more massive planets, which would then be comparable to the existence of planet d in GJ~357.

\subsection{Dynamics and TTV analysis}

While a detailed characterization of the dynamical properties of the potential planetary system is beyond the scope of this paper, we nevertheless started to investigate its properties using {\em Systemic} \citep{Meschiari2009PASP..121.1016M}. Given the fact that {\it TESS} could only cover five transits, the detection of transit timing variations (TTV) would only be possible in a system with more massive planets or in a first order resonance like, for example, in Kepler-87 \citep{2014A&A...561A.103O}. The inner pair of planets is, however, close to a 7:3 period commensurability. The dynamical interactions are small but, depending on the initial eccentricities, the system may undergo significant exchange of angular momentum, but on very long timescales of $\sim$500\,yr, which are clearly too long to be detectable in the currently available RV measurements. 

Since we cannot detect a transit for planet~c, its orbit could be inclined with respect to planet~b. The inclination of planet~c only needs to be $<88.5\pm0.1$\,deg for a non-detection, meaning a very gentle tilt with respect to planet b ($i_b = 89.12^{+0.37}_{-0.31}$\,deg) would be enough to miss it. For planet~d to transit, doing this same exercise implies one would need inclinations larger than about $89.55 +- 0.04$\,deg, which is a 1\,deg difference in mutual inclination with planet~c, and a 0.4\,deg difference with the transiting planet~b. Since multi-planet systems in general have mutual inclinations within $\sim 2$\,deg from each other \citep{Dai2018ApJ...864L..38D}, there is still room for planet d to be a transiting planet. For initially low eccentric configurations, the inclination probably cannot be constrained from dynamics, because preliminary tests show that the system is stable even for a low inclination for planet~c of $i_{\rm c}=10\,\mathrm{deg}$. At large mutual inclinations ($30\,\mathrm{deg}$), the mutual torque is significant and planet~b would drift slowly out of a transiting configuration. This effect, however, reduces significantly with a lower mutual inclination, when small mutual tilts result in torques that could bring planet~c or d into transit or planet~b out of transit. A long-term monitoring of planet b could show or constraint inclination changes from transit duration or depth variations (TDVs). 

We carried out a more in-depth search for TTVs using \textsc{PyTransit} \citep{Parviainen2015}. The approach models the near vicinity of each transit (4.8\,h around the expected transit center based on the linear ephemeris) as a product of a transit model and a flux baseline made of $n_\mathrm{L}$ Legendre polynomials. The transits are modeled jointly, and parametrized by the stellar density, impact parameter, planet-star area ratio, two quadratic limb darkening coefficients, an independent transit center for each transit, and $n_\mathrm{L}$ Legendre polynomial coefficients for each transit (modeling the baseline as a sum of polynomials rather than, for example, a Gaussian process, is still feasible given the small number of transits). The analysis results in an estimate of the model posterior distribution, where the independent parameter estimates are based on their marginal posterior distributions.
The uncertainty in the transit center estimates -- calculated as $0.5 \times \left(t_{84}-t_{16}\right)$, where $t_{16}$ and $t_{84}$ are the 16th and 84th transit center posterior percentiles -- varies from 1.5 to 4 min, and no significant deviations from the linear ephemeris can be detected, in agreement with our previous estimate.

\subsection{Formation history}

The predominant formation channel to build terrestrial planets is the core accretion scenario, where a solid core is formed before the accretion of an atmosphere sets in \citep{Mordasini2012A&A...547A.111M}. This scenario involves several stages: first, dust grows into larger particles that experience vertical settling and radial drift, commonly referred to as ``pebbles'' \citep{2012A&A...539A.148B}. They can undergo gravitational collapse into  \SI{\sim100}{\kilo\meter} sized planetesimals wherever a local concentration exceeds a level set by the disk turbulence \citep{Johansen2007, 2019ApJ...874...36L}. These planetesimals can form planetary embryos of roughly Moon size via mutual collisions \citep{2015Natur.524..322L}, at which point an accretion of further planetesimals and, even more importantly, of drifting pebbles can lead to further growth \citep{Ormel2010A&A...520A..43O}. Pebble accretion is thus very powerful in quickly forming the cores of gas giants \citep{Klahr2006, Lambrechts2014A&A...572A..35L}, but usually fails to produce terrestrial planets akin to the inner solar system due to its high efficiency. To explain the low final mass of such planets, a mechanism that stops further accretion of solids is needed. 

One way to inhibit pebble accretion is to cut the supply of solid material by another planet that forms further out earlier than or concurrently with the inner planets. Such a companion would act as a sink for the influx of material that would otherwise be available to build inner planets. This is achieved when the outer planet reaches a critical mass where the fraction of the pebble flux accreted by the planet
\begin{equation}\label{eq:Miso}
    \epsilon_{\rm PA} \approx 0.1 \times \left(\frac{q}{10^{-5}}\right)^\frac{2}{3}, \quad \mathrm{with} \, \,  q = \frac{M_\mathrm{\rm Planet}}{M_\star},
\end{equation}
approaches unity \citep{Ormel2017A&A...604A...1O}. With $q = 3 \times 10^{-5}$ and $\epsilon_{\rm PA} \approx 0.2$, GJ~357~c could have efficiently absorbed pebbles that would otherwise have reached GJ~357~b. Likewise, GJ~357~d reached a pebble accretion efficiency of $\epsilon_{\rm PA} \approx 0.3$ and could have starved the inner two planets of further accretion of pebbles. In this scenario, the planets must have formed outside-in and one would expect one or several additional planets of at least the mass of GJ~357~d further out. Such a hypothetical GJ~357~e again should have reached high pebble accretion efficiencies before its inner siblings. This is quite feasible, since the conditions for fast embryo formation are very favorable just outside the ice line of the protoplanetary disk, where the recondensation of vapor leads to a large abundance of planetesimals \citep{Stevenson1988Icar...75..146S, Cuzzi2004ApJ...614..490C, Ciesla2006Icar..181..178C, Schoonenberg2017A&A...602A..21S}. We used the minimum masses of GJ~357~c and GJ~357~d for these estimates. Thus, the efficiencies we calculated should be considered conservative, making this mechanism even more robust.

However, timing is key in order to stop the supply of solid material at the right time. The window is only open for $\sim10^5\,\mathrm{yr}$, which corresponds to the timescale for growing from a roughly Earth-sized planet to a super-Earth \citep[e.g.,][]{Bitsch2019}.

A scenario with less stringent assumptions is one where an inner planet grows to its own pebble isolation mass, which can be approximated as 
\begin{equation}\label{eq:Miso_2}
    M_\mathrm{iso} \approx h^3 M_\star
\end{equation}
with the local disk aspect ratio $h$ \citep{Ormel2017A&A...604A...1O}. Reaching $M_\mathrm{iso}$, the planet locally modifies the radial gas pressure gradient such that the inward drift of pebble-sized particles stops, starving itself and the inner system of solid material. Assuming the current orbit of GJ~357~b and an $M_\mathrm{iso}$ equal to the planetary mass inferred in our study, \autoref{eq:Miso_2} yields a local disk aspect ratio of 0.025, which is a reasonable value in the inner disk. Similarly, the minimum masses of GJ~357~c and GJ~357~d give $h \approx 0.031$ and $h \approx 0.038$, respectively, which is again consistent with estimated disk scale heights in the literature \citep[e.g.,][]{Chiang1997, Ormel2017A&A...604A...1O}.

Given the significant assumptions needed to explain the emergence of both planets by a cut-off of pebble flux in the system, the second scenario is favored. If pebble accretion is the dominating mechanism to form planetary embryos in the system, then GJ~357~b--d stopped growing when they reached their respective pebble isolation masses. However, a hypothetical future discovery of a more massive planet further out might shift the balance again towards shielding by this outer planet.

\subsection{Atmospheric characterization and habitability}

The integrated stellar flux that hits the top of an Earth-like planet’s atmosphere from a cool red star warms the planet more efficiently than the same integrated flux from a hot blue star. This is partly due to the effectiveness of the Rayleigh scattering in an atmosphere mostly composed of \ce{N2}-\ce{H2O}-CO$_2$, which decreases at longer wavelengths, together with the increased near-IR absorption by \ce{H2O} and CO$_2$.  
 
Planets GJ~357~b and GJ~357~c receive about 13 times and 4.4 times the Earth's irradiation ($S_\oplus$), respectively. Venus in comparison receives about $1.7\,S_\oplus$. Thus, both planets should have undergone a runaway greenhouse stage as proposed for Venus' evolution. Due to its incident flux level, GJ~357~c is located closer to the star than the inner edge of the empirical habitable zone as defined in \citet{Kasting1993Icar..101..108K} and \citet{Kopparapu14}. On the other hand, GJ~357~d receives an irradiation of $0.38\,S_\oplus$ , which places it inside the habitable zone (as defined above), in a location comparable to Mars in the solar sytem, making it a very interesting target for further atmospheric observations.  

Atmospheric characterization of exoplanets is difficult because of the high contrast ratio between a planet and its host star. While atmospheres of Earth and super-Earth planets are still outside our technical capabilities, upcoming space missions such as the {\it James Webb Space Telescope (JWST)} and the extremely large ground-based telescopes (ELTs) will open this possibility for a selected group of rocky planets offering the most favorable conditions. 

\begin{figure}
\centering
\includegraphics[trim=1 3 2 5, clip, width=\hsize]{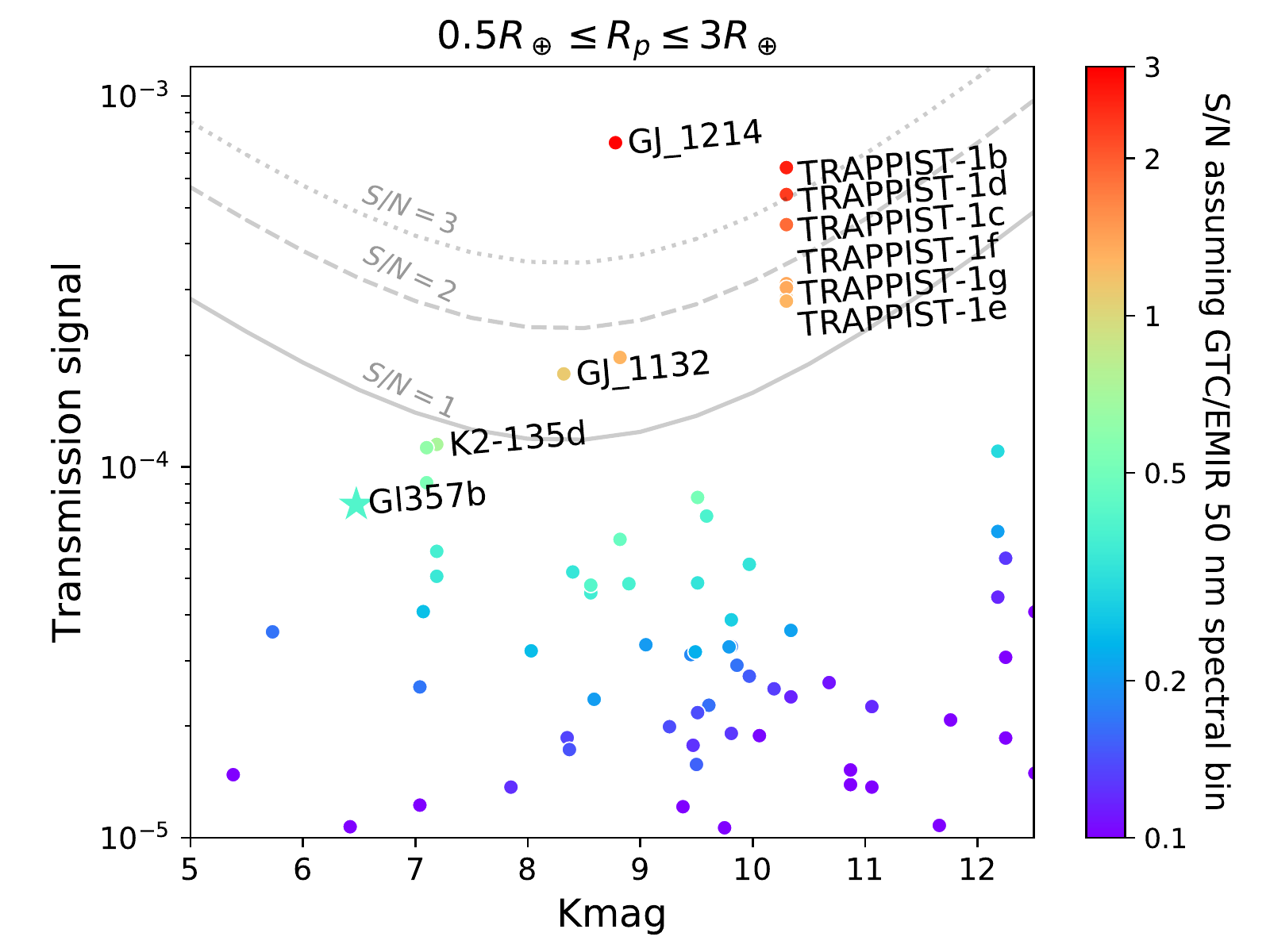}
\caption{Expected primary transit transmission signal per scale height plotted against the $K$-band magnitude of GJ~357~b (star) and all know planets (dots) with mass measurements and with radius between 0.5 and 3\,$R_\oplus$.  The color scale provides the expected S/N for a single transit assuming the use of a ten-meter telescope and 50\,nm wavelength integration bins. The gray lines indicate the pattern of the S/N assuming a single transit duration of 3\,h. The relative S/N is maintained when extrapolated to other instrumentation. A few benchmark targets for {\it JWST} transmission spectroscopy studies are labeled.}
\label{fig:trans}
\end{figure}

The star GJ~357 is one of the brightest M dwarfs in the sky, and as such, planets orbiting it are interesting targets for follow-up characterization. To illustrate this fact, in Fig.~\ref{fig:trans} we plot the expected transmission signal reachable in a single transit with a ground-based 10\,m telescope for all known planets with mass measurements and with radius between 0.5 and 3\,$R_\oplus$. { The transmission signal per scale height is defined as
\begin{equation}
  TS = 2 H_s \frac{R_p}{R_s^2}
,\end{equation}
where $R_p$ and $R_s$ are the radii of planet and star, respectively, and $H$ is the scale height   
\begin{equation}
  H_s = \frac{k_{\rm B} T_{\rm eq}}{\mu g_p}  
,\end{equation}
where $k_{\rm B}$ is the Boltzmann constant, $T_{\rm eq}$ and $g_p$ are the equilibrium temperature and surface gravity of the planet, respectively, and $\mu=2.3\,\mathrm{g\,mol^{-1}}$ is the mean molecular weight. The signal is calculated then as $1.8\times TS$ where a spectral modulation of $1.8\,H_s$ is adopted \citep{2016ApJ...823..109I}. This signal is an optimistic estimate, because terrestrial planets are unlikely to host an atmosphere of mean molecular weight at $2.3\,\mathrm{g\,mol^{-1}}$.} The most favorable planets for atmospheric characterization offer a combination of a large scale height (puffiness of the atmospheres) and host star brightness, and are labeled in the figure together with GJ~357~b. 

\citet{2018PASP..130k4401K} proposed a metric to select {\it TESS} (and other missions) planet candidates according to their suitability for atmospheric characterization studies. Using the mass and radius determined in this work ($1.84\,M_\oplus$, $1.217\,R_\oplus$), we obtained a transmission metric value of 23.4 for GJ~357~b. For comparison, two of the most well-known planets around bright M-type stars with favorable metrics, LHS~1140~b and TRAPPIST-1~f, have metric values of 9.13 and 13.7, respectively. It is worth noting that out of the simulated yield of {\it TESS} terrestrial planets with $R<2\,R_\oplus$ used in \citet{2018PASP..130k4401K}, in turn based on \citet{2015ApJ...809...77S} and assuming an Earth-like composition, only one had a larger metric value (28.2). Using the same reference, the emission spectroscopy metric for GJ~357~b is 4.1, a modest number compared to the simulated yield of TESS planets suitable for these types of studies. 

\begin{figure}
    \centering
    \includegraphics[width=\hsize]{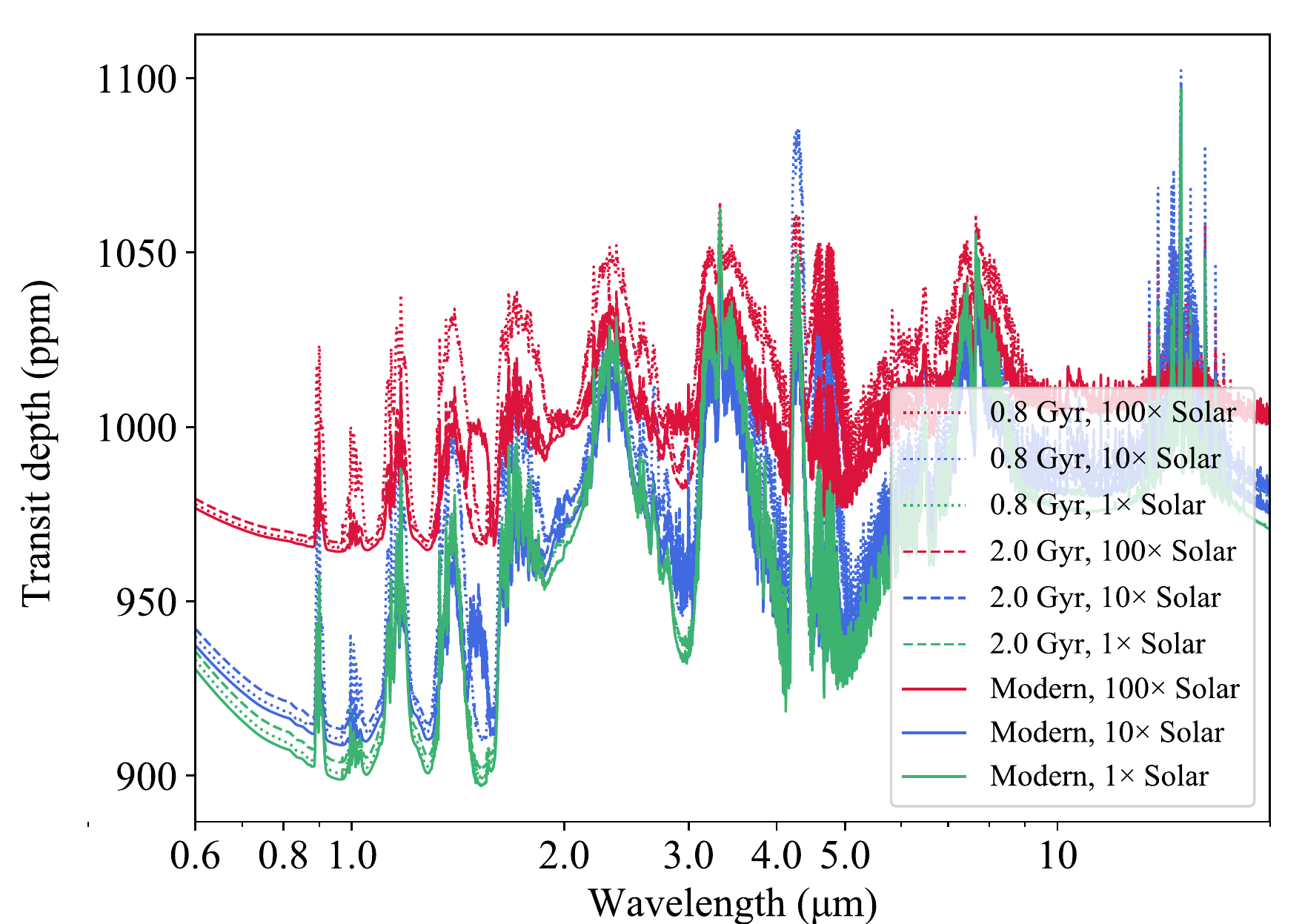}
    \caption{Synthetic spectra of GJ~357~b in the {\it JWST} wavelength range. Nine transmission spectra, including three atmospheric metallicities and three temperature structures from three different Earth epochs are shown. The assumed temperature profile is the Earth's one adapted for $T_\mathrm{eff}=525$\,K to resemble a hot-terminator scenario.} 
    \label{fig:synspec_b}
\end{figure}

In order to assess an estimation of GJ~357~b's atmospheric signal through transmission spectroscopy, we simulated a simplified atmospheric photochemistry model for a rocky planet basing it on early Earth's temperature structure, increasing the surface temperature to be consistent with $T_\mathrm{eff}=525$\,K  and removing water from the atmosphere using \texttt{ChemKM} (Molaverdikhani in prep.). The temperature and pressure structures of GJ~357~b's atmosphere are not modeled self consistently, as this only shows sample spectra.

Geometric mean spectra of GJ~667~C ($T_\mathrm{eff}=3327\,\mathrm{K}$) and GJ~832 ($T_\mathrm{eff}=3816$\,K) were considered as an estimation of GJ~357's flux in the range of X-ray to optical wavelengths. The data were obtained from the MUSCLES database \citep{france2016muscles}. We modeled three different metallicities, 1$\times$, 10$\times$, and 100$\times$ solar metallicity to explore a wider range of possibilities \citep{wakeford2017hat} and selected a temperature and atmosphere profile based on an anoxic Earth atmosphere. We selected three geological epochs, namely 2.0\,Gyr (after the Great Oxygenation Event), 0.8\,Gyr (after the Neoproterozoic Oxygenation Event, when multicellular life began to emerge), and the modern Earth \citep{kawashima2019theoretical}, to consider three different atmospheric conditions with different temperature structures.  To set up the models, we used \citet{venot2012chemical}'s full kinetic network and an updated version of \citet{hebrard2012neutral}'s UV absorption cross sections and branching yields. 

Synthetic sample transmission spectra for GJ~357~b are calculated using {\tt petitRADTRANS} \citep{petitRADTRANS}, shown in Fig.~\ref{fig:synspec_b}. The major opacity source in the atmosphere is mostly methane, and as expected CH$_4$ and CO$_2$ contribute more significantly at higher temperatures and metallicities in this class of planets \citep{molaverdikhani2019cold}. Such spectral features are expected to be above {\it JWST}'s noise floor; 20\,ppm, 30\,ppm, and 50\,ppm, for NIRISS SOSS, NIRCam grism, and MIRI LRS, respectively \citep{greene2016characterizing}. Ground-based, high-resolution spectroscopy can potentially access these strong absorption lines too. We must emphasize that these synthetic spectra are calculated under the assumption of cloud-free atmospheres. If clouds were present, the spectral features could be obscured or completely muted, resulting in a flattened transmission spectrum \citep{kreidberg2014clouds}.

As a final note on the atmospheric characterization of GJ~357~b, the star may be too bright in some of the {\it JWST}'s observing modes, which demands careful observational planning for transmission spectroscopy. Such bright objects, however, are excellent for ground-based facilities. \citet{2018PASP..130d4401L} pointed out that only a few {\it TESS} planets of terrestrial size are expected to be as good or better targets for atmospheric characterization than the currently known planets.  GJ~357~b is one of them and, although it is not in the habitable zone, in fact it could be so far the best terrestrial planet for atmospheric characterization with the upcoming {\it JWST} and ground-based ELTs. 

\begin{figure}
    \centering
    \includegraphics[trim=0 0 0 5, clip,width=\hsize]{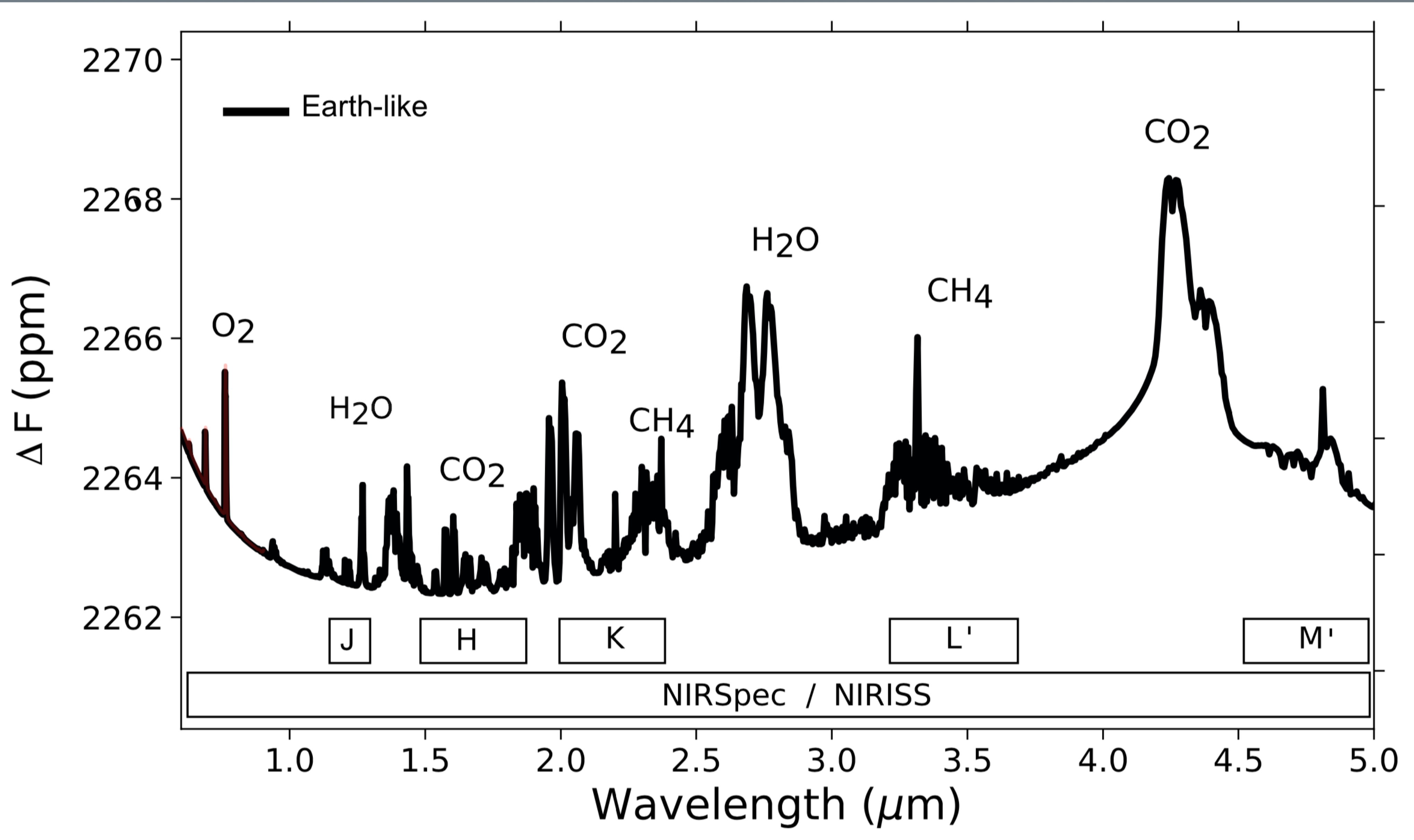} \\
    \includegraphics[trim=0 0 0 5, clip,width=\hsize]{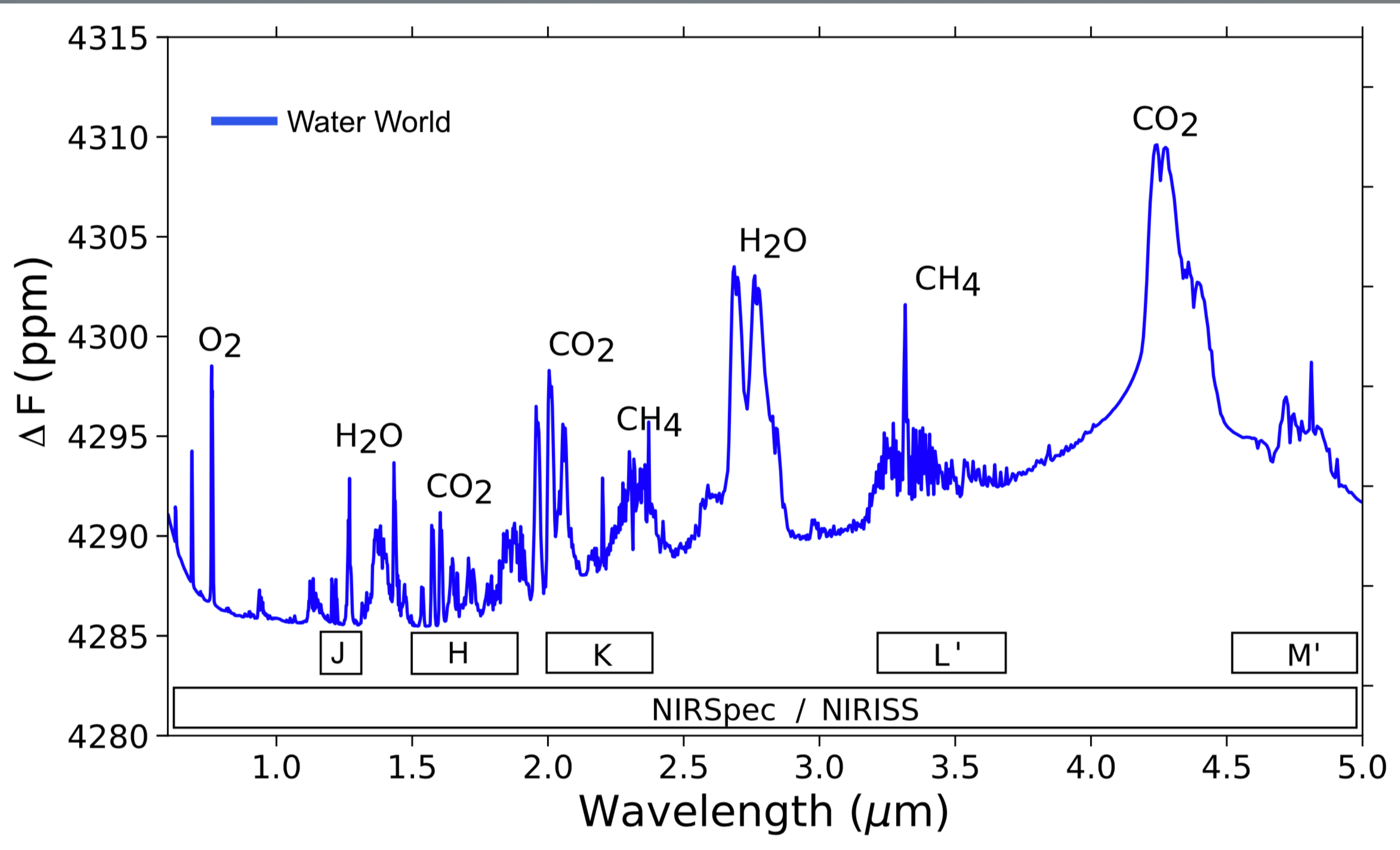}
    \caption{Synthetic spectra for GJ~357~d in the {\it JWST} NIRISS/NIRSpec wavelength range for two models: Earth-like composition and atmosphere (assuming a radius of $1.75\,R_\oplus$, black) and water world composition with Earth-like atmosphere (assuming a radius of $2.4\,R_\oplus$,blue) as an example for detectable features.} 
    \label{fig:synspec_d}
\end{figure}

For GJ~357~d, a rocky Earth-like composition corresponds to a $1.75\,R_\oplus$ planet, while an ice composition would correspond to a planet radius of $2.4\,R_\oplus$. We still do not know whether GJ~357~d transits its host star, however, if it did, the atmospheric signal (assuming an Earth-like composition and atmosphere) would become detectable by {\it JWST} for both NIRISS/NIRSpec (Fig.~\ref{fig:synspec_d}) and MIRI. Self-consistent models of the planet, as well as an expected atmospheric signal for GJ~357~d, assuming a range of different compositions and atmospheres, show cool surface temperatures for Earth-like models and warm conditions for early Earth-like models. The models as well as the observable spectral features have been generated using EXO-Prime \citep[see, e.g.,][]{Kaltenegger2010ApJ...708.1162K} 
and are discussed in detail in  Kaltenegger et al. (2019, in prep.). The code incorporates a 1D climate, 1D photochemistry, and 1D radiative transfer model that simulates both the effects of stellar radiation on a planetary environment and the planet's outgoing spectrum.

\section{Conclusions} \label{sec:conclusions}

We report the discovery and confirmation of a planetary system around the bright M dwarf GJ~357. Data from the {\it TESS} mission revealed the first clue of this discovery by detecting the transit signals of GJ~357~b through the {\it TESS} Alerts website. The availability of archival and new high-precision RV data made possible the quick confirmation of GJ~357~b, and a search for further planet candidates in the system. 

Planet GJ~357~b is a hot Earth-sized transiting planet with a mass of $1.84\pm0.31\,M_\oplus$ in a 3.93\,d orbit. The brightness of the planet's host star makes GJ~357~b one of the prime targets for future atmospheric characterization, and arguably one of the best future opportunities to characterize a terrestrial planet atmosphere with {\it JWST} and ELTs. To date, GJ~357~b is the nearest transiting planet to the Sun around an M dwarf and contributes to the {\it TESS} Level One Science Requirement of delivering 50 transiting small planets (with radii smaller than $4\,R_\oplus$) with measured masses to the community.

Finally, there is evidence for at least two more planets, namely GJ~357~c, with a minimum mass of $3.4\pm0.46\,M_\oplus$ in a 9.12\,d orbit, and GJ~357~d, an interesting super-Earth or sub-Neptune with a minimum mass of $6.1\pm1.0\,M_\oplus$ in a 55.7\,d orbit inside the habitable zone. Thus, GJ~357 adds to the growing list of {\it TESS} discoveries deserving more in-depth studies, as these systems can provide relevant information for our understanding of planet formation and evolution.

\begin{acknowledgements}

This paper includes data collected by the \textit{TESS} mission. Funding for the \textit{TESS} mission is provided by the NASA Explorer Program. We acknowledge the use of \textit{TESS} Alert data, which is currently in a beta test phase, from pipelines at the \textit{TESS} Science Office and at the \textit{TESS} Science Processing Operations Center. Resources supporting this work were provided by the NASA High-End Computing (HEC) Program through the NASA Advanced Supercomputing (NAS) Division at Ames Research Center for the production of the SPOC data products. This research has made use of the Exoplanet Follow-up Observation Program website, which is operated by the California Institute of Technology, under contract with the National Aeronautics and Space Administration under the Exoplanet Exploration Program. 

This work has made use of data from the European Space Agency (ESA) mission {\it Gaia} (\url{https://www.cosmos.esa.int/gaia}), processed by the {\it Gaia} Data Processing and Analysis Consortium (DPAC, \url{https://www.cosmos.esa.int/web/gaia/dpac/consortium}). Funding for the DPAC has been provided by national institutions, in particular the institutions participating in the {\it Gaia} Multilateral Agreement.

CARMENES is an instrument for the Centro Astron\'omico Hispano-Alem\'an de Calar Alto (CAHA, Almer\'ia, Spain) funded by the German Max-Planck-Gesellschaft (MPG), the Spanish Consejo Superior de Investigaciones Cient\'ificas (CSIC), the European Union through FEDER/ERF~FICTS-2011-02 funds, and the members of the CARMENES Consortium.

R.\,L. has received funding from the European Union’s Horizon 2020 research and innovation program under the Marie Skłodowska-Curie grant agreement No.~713673 and financial support through the “la Caixa” INPhINIT Fellowship Grant LCF/BQ/IN17/11620033 for Doctoral studies at Spanish Research Centers of Excellence from “la Caixa” Banking Foundation, Barcelona, Spain. This work is partly financed by the Spanish Ministry of Economics and Competitiveness through projects ESP2016-80435-C2-2-R and ESP2016-80435-C2-1-R.

We acknowledge support from the Deutsche Forschungsgemeinschaft under DFG Research Unit FOR2544 ``Blue Planets around Red Stars'', project no.\ QU 113/4-1, QU 113/5-1, RE 1664/14-1, DR 281/32-1, JE 701/3-1, RE 2694/4-1, the Klaus Tschira Foundation, and the Heising Simons Foundation.
This work is partly supported by JSPS KAKENHI Grant Numbers JP15H02063, JP18H01265, JP18H05439, JP18H05442, and JST PRESTO Grant Number JPMJPR1775. 

This research has made use of the services of the ESO Science Archive Facility. Based on observations collected at the European Southern Observatory under ESO programs 
072.C-0488(E), 183.C-0437(A), 
072.C-0495, 078.C-0829, and 173.C-0606. 
IRD is operated by the Astrobiology Center of the National Institutes of Natural Sciences. 
M.S.\ thanks Bertram Bitsch for stimulating discussions about pebble accretion.
\end{acknowledgements}

\bibliographystyle{aa} 
\bibliography{biblio} 


\begin{appendix} 

\section{Joint fit priors.}

\begin{table*}
    \centering
    \caption{Priors used for the joint fit model 3pl+GPexp presented in \autoref{subsubsec:joint} using \texttt{juliet}. The prior labels of $\mathcal{N}$, $\mathcal{U}$, and $\mathcal{J}$ represent normal, uniform, and Jeffrey's distributions. The parametrization for $(p,b)$ using $(r_1,r_2)$ \citep{Espinoza18} and the linear $(q_1)$ and quadratic $(q_1,q_2)$ limb-darkening parametrization \citep{Kipping13} are both described in \autoref{subsubsec:photonly}.}
    \label{tab:priors}
    \begin{tabular}{lccr} 
        \hline
        \hline
        \noalign{\smallskip}
        Parameter name & Prior & Units & Description \\
        \noalign{\smallskip}
        \hline
        \noalign{\smallskip}
        \multicolumn{4}{c}{\it Stellar parameters} \\
        \noalign{\smallskip}
        $\rho_\star$ & $\mathcal{N}(13600,1700^2)$ & $\mathrm{kg\,m\,^{-3}}$ & Stellar density. \\
        \noalign{\smallskip}
        \multicolumn{4}{c}{\it Planet parameters} \\
        \noalign{\smallskip}
        $P_{\rm b}$              & $\mathcal{N}(3.93079,0.001^2)$    & d                 & Period of planet b. \\
        $P_{\rm c}$              & $\mathcal{N}(9.1,0.1^2)$          & d                 & Period of planet c. \\
        $P_{\rm d}$              & $\mathcal{N}(55.7,0.5^2)$          & d                 & Period of planet d. \\
        $t_{0,b} - 2450000$      & $\mathcal{N}(8517.99,0.1^2)$    & d                 & Time of transit-center of planet b. \\
        $t_{0,c} - 2450000$      & $\mathcal{U}(8312,8318)$    & d                 & Time of transit-center of planet c. \\
        $t_{0,d} - 2450000$      & $\mathcal{U}(8310,8340)$    & d                 & Time of transit-center of planet d. \\
        $r_{1,b}$                & $\mathcal{U}(0,1)$            & \dots             & Parametrization for $p$ and $b$. \\
        $r_{2,b}$                & $\mathcal{U}(0,1)$            & \dots             & Parametrization for $p$ and $b$. \\
        $K_{b}$                  & $\mathcal{U}(0,10)$          & $\mathrm{m\,s^{-1}}$     & RV semi-amplitude of planet b. \\
        $K_{c}$                  & $\mathcal{U}(0,10)$          & $\mathrm{m\,s^{-1}}$     & RV semi-amplitude of planet c. \\
        $K_{d}$                  & $\mathcal{U}(0,10)$          & $\mathrm{m\,s^{-1}}$     & RV semi-amplitude of planet d. \\
        $\mathcal{S}_{1,b,c,d} = \sqrt{e_{b,c,d}}\sin \omega_{b,c,d}$    & 0.0 (fixed) & \dots  & Parametrization for $e$ and $\omega$. \\
        $\mathcal{S}_{2,b,c,d} = \sqrt{e_{b,c,d}}\cos \omega_{b,c,d}$    & 0.0 (fixed) & \dots  & Parametrization for $e$ and $\omega$. \\
        \noalign{\smallskip}
        \multicolumn{4}{c}{\it Photometry parameters} \\
        \noalign{\smallskip}
        $D_{\textnormal{TESS}}$                  & 1.0 (fixed)               & \dots     & Dilution factor. \\
        $M_{\textnormal{TESS}}$                  & $\mathcal{N}(0,0.1^2)$    & ppm       & Relative flux offset for {\it TESS}. \\
        $\sigma_{\textnormal{TESS}}$           & $\mathcal{U}(1,500)$      & ppm       & Extra jitter term for {\it TESS}. \\
        $q_{1,\textnormal{TESS}}$                & $\mathcal{U}(0,1)$        & \dots     & Quadratic limb-darkening parametrization for {\it TESS}. \\
        $q_{2,\textnormal{TESS}}$                & $\mathcal{U}(0,1)$        & \dots     & Quadratic limb-darkening parametrization for {\it TESS}. \\
        $M_{\textnormal{LCO}}$                  & $\mathcal{N}(0,0.1^2)$    & ppm       & Relative flux offset for LCO. \\
        $\sigma_{\textnormal{LCO}}$           & $\mathcal{U}(1,2000)$      & ppm       & Extra jitter term for LCO. \\
        $q_{1,\textnormal{LCO}}$                & $\mathcal{U}(0,1)$        & \dots     & Linear limb-darkening parametrization for LCO. \\
        \noalign{\smallskip}
        \multicolumn{4}{c}{\it RV parameters} \\
        \noalign{\smallskip}
        $\mu_{\textnormal{HIRES}}$           & $\mathcal{U}(-10,10)$    & $\mathrm{m\,s^{-1}}$ & Systemic velocity for HIRES. \\
        $\sigma_{\textnormal{HIRES}}$      & $\mathcal{U}(0,10)$      & $\mathrm{m\,s^{-1}}$ & Extra jitter term for HIRES. \\
        $\mu_{\textnormal{UVES}}$            & $\mathcal{U}(-10,10)$    & $\mathrm{m\,s^{-1}}$ & Systemic velocity for UVES. \\
        $\sigma_{\textnormal{UVES}}$       & $\mathcal{U}(0,10)$      & $\mathrm{m\,s^{-1}}$ & Extra jitter term for UVES. \\
        $\mu_{\textnormal{HARPS}}$           & $\mathcal{U}(-10,10)$    & $\mathrm{m\,s^{-1}}$ & Systemic velocity for HARPS. \\
        $\sigma_{\textnormal{HARPS}}$      & $\mathcal{U}(0,10)$      & $\mathrm{m\,s^{-1}}$ & Extra jitter term for HARPS. \\
        $\mu_{\textnormal{PFSpre}}$           & $\mathcal{U}(-10,10)$    & $\mathrm{m\,s^{-1}}$ & Systemic velocity for PFSpre. \\
        $\sigma_{\textnormal{PFSpre}}$      & $\mathcal{U}(0,10)$      & $\mathrm{m\,s^{-1}}$ & Extra jitter term for PFSpre. \\
        $\mu_{\textnormal{PFSpost}}$           & $\mathcal{U}(-10,10)$    & $\mathrm{m\,s^{-1}}$ & Systemic velocity for PFSpost. \\
        $\sigma_{\textnormal{PFSpost}}$      & $\mathcal{U}(0,10)$      & $\mathrm{m\,s^{-1}}$ & Extra jitter term for PFSpost. \\
        $\mu_{\textnormal{CARMENES}}$        & $\mathcal{U}(-10,10)$    & $\mathrm{m\,s^{-1}}$ & Systemic velocity for CARMENES. \\
        $\sigma_{\textnormal{CARMENES}}$   & $\mathcal{U}(0,10)$      & $\mathrm{m\,s^{-1}}$ & Extra jitter term for CARMENES. \\
        \noalign{\smallskip}
        \noalign{\smallskip}
        \multicolumn{4}{c}{\it GP hyperparameters} \\
        \noalign{\smallskip}
        $\sigma_\mathrm{GP,TESS}$            & $\mathcal{J}(10^{-2},500)$     & ppm                           & Amplitude of GP component for TESS. \\
        $T_\mathrm{GP,TESS}$            & $\mathcal{J}(10^{-2},1000)$     & d                        & Length scale of GP component for TESS. \\
        $\sigma_\mathrm{GP,RV}$              & $\mathcal{J}(0.1,10)$          & $\mathrm{m\,s^{-1}}$          & Amplitude of GP component for the RVs. \\
        $T_\mathrm{GP,RV}$              & $\mathcal{J}(10^{-4},10)$     & d                        & Length scale of GP component for the RVs. \\
        \noalign{\smallskip}
        \hline
    \end{tabular}
\end{table*}

\clearpage

\section{RV time series of best joint fit}

\begin{figure*}
\centering
\includegraphics[width=\hsize]{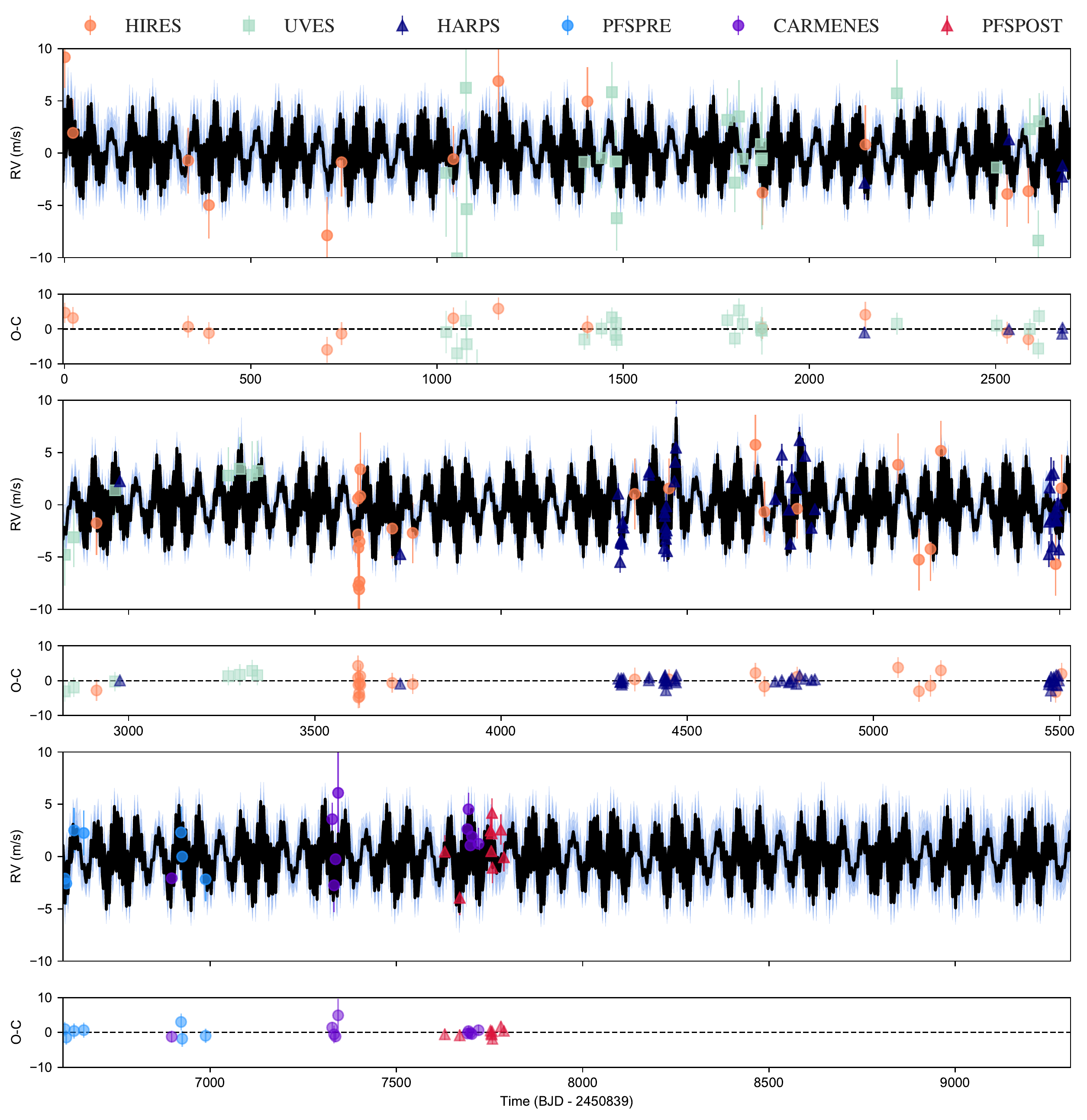}
\caption{RV measurements as a function of time along with the residuals obtained from subtracting our median best joint fit model (black line) and the 68\%, 95\%, and 99\% posterior bands (shown in blue). The color coding of the datapoints for each instrument is shown at the top.} 
\label{fig:rv_timeseries1}
\end{figure*}

\end{appendix}

\end{document}